\documentclass[twocolumn,tighten]{aastex631}
\usepackage{amsfonts,amsmath,amssymb}
\usepackage{bm}
\usepackage{xcolor}
\usepackage[]{cleveref}
\crefname{section}{\S\!}{\S\S\!}
\crefname{appendix}{\S\!}{\S\S\!}
\crefname{equation}{Eq.}{Eqs.}
\Crefname{equation}{Equation}{Equations}
\crefname{figure}{Fig.}{Figs.}
\Crefname{figure}{Figure}{Figures}
\usepackage{mathtools,upgreek,pifont}
\usepackage{cuted}

\def\apj{ApJ}
\def\apjl{ApJL}

\def\aap{A\&A}
\def\rvgeo{RvGeo}

\def\mnras{MNRAS}
\def\nat{Natur}
\def\pnas{PNAS}

\def\prl{PhRvL}
\def\prx{PhRvX}
\def\jgr{JGR}
\def\jgra{JGRA}

\def\ssr{SSRv}

\def\pop{PhPl}
\def\pof{PhFl}

\def\jcomp{JCoPh}
\def\jpp{JPlPh}

\def\lrvsp{LRSP}

\def\solphys{SoPh}
\def\aapr{A\&ARv}
\def\natas{NatAs}

\def\jcomp{JCoPh}

\def\jetp{JETP}
\def\jatp{JATP}

\usepackage{hyperref}
\definecolor{darkblue}{rgb}{0.0,0.0,0.3}
\hypersetup{colorlinks,breaklinks,
            linkcolor=darkblue,urlcolor=darkblue,
            anchorcolor=darkblue,citecolor=darkblue}

\usepackage{graphicx}
\usepackage[utf8]{inputenc}
\usepackage{cancel}
\usepackage{commath}
\usepackage{dsfont}
\usepackage{xfrac}
\usepackage{xparse}
\newcommand\bb[1]{\mbox{\boldmath{$#1$}}}
\newcommand\grad{\bb{\nabla}}
\newcommand\bcdot{\,\bb{\cdot}\,}

\newcommand\btimes{\,\bb{\times}\,}

\newcommand\bs[1]{\boldsymbol{#1}}

\newcommand{\imag}{{\rm i}}
\newcommand{\const}{{\rm const}}

\newcommand{\rmd}{{\rm d}}
\newcommand{\rme}{{\rm e}}

\newcommand{\vth}[1]{v_{{\rm th}{#1}}}
\newcommand{\rmJ}{{\rm J}}

\ifcsname pdfoutput\endcsname
  \ifnum\pdfoutput>0 
    \newsavebox\diffdbox{}
    \newcommand{\slantedromand}{{\mathpalette\makesl{d}}}
    
  \else 
    \newcommand{\slantedromand}{d} 
  \fi
\else 
  \newcommand{\slantedromand}{d} 
\fi
\makeatletter
\newcommand{\dd}[1][]{\mathop{}\!%
  \expandafter\ifx\expandafter&\detokenize{#1}&
    \slantedromand\@ifnextchar^{\hspace{0.2ex}}{\hspace{0.1ex}}
  \else
    \slantedromand\hspace{0.2ex}^{#1}
  \fi}
\makeatother
\NewCommandCopy{\daccent}{\d}
\renewcommand{\d}{\ifmmode\dd\else\daccent\fi}

\ProvideDocumentCommand\dv{o m g}{%
  \IfNoValueTF{#3}{%
    \dv[#1]{}{#2}}{%
    \IfNoValueTF{#1}{%
      \frac{\dd #2}{\dd #3}%
    }{\frac{\dd[#1] #2}{\dd {#3}^{#1}}}}}
\ExplSyntaxOn
\ProvideDocumentCommand\pdv{o m g}{%
  \IfNoValueTF{#3}{\pdv[#1]{}{#2}}%
  {\ifnum\clist_count:n{#3}<2
      \IfValueTF{#1}{\frac{\partial^{#1} #2}{\partial {#3}^{#1}}}%
      {\frac{\partial #2}{\partial #3}}
    \else
      \frac{\IfValueTF{#1}{\partial^{#1}}{\partial^{\clist_count:n{#3}}}#2}%
      {\clist_map_inline:nn{#3}{\partial ##1 \,}\!}
    \fi}}
\ExplSyntaxOff

\shorttitle{Inverted minor-ion heating in imbalanced turbulence}
\shortauthors{Zhang et al.}

\begin{document}

\title[]{Minor Ions as a Diagnostic of Solar Wind Heating: \\ Inverted Mass-to-Charge Scaling in Imbalanced Turbulence}
\author[0000-0002-3987-5977]{Michael F.~Zhang}
\affiliation{Physics Department, University of Otago, 730 Cumberland St, Dunedin 9016, New Zealand}
\affiliation{Department of Astrophysical Sciences, 
Princeton University, Peyton Hall, Princeton, NJ 08544, USA}
\affiliation{Princeton Plasma Physics Laboratory, PO~Box 451, Princeton, NJ 08543, USA}
\email{E-mail for correspondence: mzhang@otago.ac.nz}
\author[0000-0002-9348-1290]{Evan L.~Yerger}
\affiliation{Space Science Center and Department of Physics and Astronomy, University of New Hampshire,
Durham, NH 03824, USA}
\author[0000-0003-1676-6126]{Matthew W.~Kunz}
\affiliation{Department of Astrophysical Sciences, 
Princeton University, Peyton Hall, Princeton, NJ 08544, USA}
\affiliation{Princeton Plasma Physics Laboratory, PO~Box 451, Princeton, NJ 08543, USA}
\author[0000-0001-8479-962X]{Jonathan Squire}
\affiliation{Physics Department, University of Otago, 730 Cumberland St, Dunedin 9016, New Zealand}

\begin{abstract}

Alfv\'enic turbulence is thought to be vital to powering the solar wind and corona, yet has eluded a comprehensive understanding of the kinetic processes by which it dissipates. Minor ions serve as sensitive tracers of these processes, showing extreme perpendicular temperatures relative to the local magnetic field and, perplexingly, mass-weighted temperature trends that sometimes correlate, but sometimes anticorrelate, with mass-to-charge ratio, \(A_i/Z_i\). We use a combination of quasilinear theory and 3D hybrid-kinetic particle-in-cell simulations to explain these features and predict further correlations with other properties of the turbulence in the fast solar wind. When Alfv\'enic turbulence is imbalanced, its cascade to ion-Larmor scales is throttled by the helicity barrier. This barrier ultimately leads to high-frequency proton-cyclotron waves (PCWs), both oblique and parallel, the latter of which produce very flat electric-energy spectra 
(\(\mathcal{E}_E \sim k_\parallel^{-\eta}\) with \(\eta<2\)) over the range of scales that are cyclotron resonant with minor ions. While steeper spectra lead to a positive correlation of heating with \(A_i/Z_i\), the shallower spectra cause the dependence to invert, with \(Q_i \propto Q_{\mathrm{p}} A_i(A_i/Z_i)^{\eta-2} \). This result is corroborated by a set of six simulations of both balanced and imbalanced turbulence, spanning \(\beta_{\rm p0} = \{1, 0.3, 1/16\}\), which demonstrate minor-ion heating rates following the power-law scaling \((A_{i} / Z_{i})^a\). We show that minor-ion heating is strongest and most perpendicular in imbalanced turbulence at lower \(\beta_{\rm p0}\), with extreme temperature ratios \(T_{\perp{\rm O^{5+}}}/T_{\perp {\rm p}} \approx 40\) and anisotropic \(T_{\perp{\rm O}^{5+}}/T_{\parallel{\rm O}^{5+}} \sim 10\) at \(\beta_{\rm p0} = 1/16\), in agreement with low-coronal observations of extreme temperature ratios and anisotropies. Future minor-ion measurements should test whether intervals in which minor-ion thermal speeds decrease with increasing mass-to-charge ratio are associated with a history of large cross helicity, enhanced power in parallel PCWs, and a steep transition-range spectrum.

\end{abstract}

\keywords{Solar wind (1534); Solar coronal heating (1989); Space plasmas (1544); Interplanetary turbulence (830); Plasma astrophysics (1261)}

\section{Introduction}
\label{sec:introduction}

Both the solar corona and the solar wind require substantial heating far above the surface of the Sun. Although the mechanisms responsible for this heating remain uncertain, the required energy likely originates from Alfv\'enic turbulence, which observations suggest is energetically sufficient to power the fast wind \citep{mcintoshAlfvenicWavesSufficient2011,halekasQuantifyingEnergyBudget2023,riveraSituObservationsLargeamplitude2024}.
In the weakly collisional solar wind, particle species need not remain in thermal equilibrium; thus, the dissipation of turbulent energy---and its partition between species and between directions parallel and perpendicular to the local magnetic field---is central to determining fast-wind properties \citep{verscharenMultiscaleNatureSolar2019}.
Minor ions are especially sensitive probes of this physics. Their wide range of masses and charge states makes their thermodynamic properties stringent tests of theories for turbulent dissipation and wave--particle interactions in collisionless plasmas \citep{bochslerMinorIonsSolar2007,vonsteigerCompositionQuasistationarySolar2000,bochslerKineticTemperaturesHeavy1985}. They trace these processes throughout the inner heliosphere, down to coronal heights where charge states freeze in and ion species thermally decouple \citep{bameQuietCoronaTemperature1974,koEmpiricalStudyElectron1997}.

Minor ions also exhibit several striking observational signatures.
Their temperatures often greatly exceed proton temperatures, especially perpendicular to the local magnetic-field direction.
Remote UVCS observations of coronal holes, for example, infer perpendicular \(\mathrm{O}^{5+}\) temperatures more than \(40\) times higher than proton temperatures, \(T_{\perp\mathrm{O}^{5+}}/T_{\perp\mathrm{p}}\gtrsim40\), together with magnetic-field-biased temperature anisotropies as large as \(T_{\perp\mathrm{O}^{5+}}/T_{\parallel\mathrm{O}^{5+}}\sim10\)
\citep{kohlUltravioletSpectroscopyExtended2006,cranmerImprovedConstraintsPreferential2008,cranmerCoronalHoles2009a}.
More recently, Solar Orbiter measurements beyond \(0.3\)~au find \(\mathrm{O}^{6+}\) temperatures exceeding proton temperatures by more than a factor of \(30\), \(T_{\mathrm{O}^{6+}}/T_{\mathrm{p}}\gtrsim30\) \citep{riveraObservationalConstraintsRadial2025}.
Likewise, near-Sun in-situ measurements from Parker Solar Probe find alpha temperatures \(6{-}16\) times those of protons \citep{mostafaviParkerSolarProbe2024a}, with preferentially perpendicular heating resulting in  \(T_{\perp\alpha}/T_{\parallel\alpha}\sim3\)
\citep{mostafaviPreferentialEnergizationSolar2025}.

\begin{figure}[t]
    \centering
    \includegraphics[width=0.95\columnwidth]{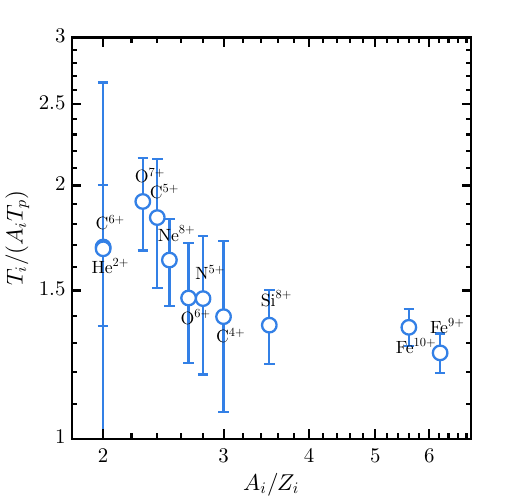}
    \caption{Minor-ion to proton temperature ratios, normalized by the ion-to-proton mass ratio \(A_i\), plotted against ion-to-proton mass-to-charge ratio \(A_i/Z_i\).
Data correspond to \emph{ACE}/\emph{SWICS} measurements at \(1\)~au for a collisionally young solar-wind interval, adapted from fig.~3 of \citet{tracyConstrainingSolarWind2016}.}
    \label{fig:tracy2016}
\end{figure}

Even more puzzling is the observed dependence of minor-ion temperatures on mass-to-charge ratio, \(A_i/Z_i\), where \(A_i\) and \(Z_i\) are the ion-to-proton mass and charge ratios,
\begin{equation}\label{eq:aizi}
    A_i \doteq \frac{m_i}{m_{\rm p}}, \qquad Z_i \doteq \frac{q_i}{q_{\rm p}} .
    \refstepcounter{equation}\tag{\theequation a,\,b}
\end{equation}
Depending on the range of \(A_i/Z_i\) considered, minor-ion temperatures can correlate positively or negatively with \(A_i/Z_i\).
Fig.~\ref{fig:tracy2016} shows an example of negatively correlated temperatures in collisionally young solar wind at \(1\)~au \citep{tracyConstrainingSolarWind2016}.
Similar features are also observed in polar coronal holes \citep{landiIonTemperaturesLow2009a}. 
When minor-ion temperatures decrease with \(A_i/Z_i\), as for \(A_i/Z_i\in[2.2,3]\) in Fig.~\ref{fig:tracy2016}, we refer to the scaling as ``inverted.''
The goal of this paper is to explain how such correlations can arise and change with the properties of the turbulence, thereby advancing the broader use of minor ions as detailed diagnostics of coronal and solar-wind heating.

We show that these correlations, together with the other key features of minor-ion observations discussed above, arise naturally from heating by Alfv\'enic turbulence when cross helicity---the energy imbalance between outward- and inward-propagating fluctuations in the solar wind---is taken into account.
To explain inverted \(A_i/Z_i\) trends in particular, we develop a phenomenology, rooted in quasilinear theory, in which the dominant heating channel changes as ion velocity distribution functions (VDFs) approach saturation and parallel proton-cyclotron waves (PCWs) emerge self-consistently.
The resulting competition between oblique Alfv\'enic fluctuations and coherent parallel PCWs determines whether heating increases or decreases with mass-to-charge ratio, \(A_i/Z_i\).
We test these predictions against 3D hybrid-kinetic simulations of minor-ion heating in both balanced and imbalanced turbulence.

This focus on turbulence imbalance reflects its key role, alongside the ratio of thermal to magnetic pressure (plasma beta \(\beta\)), in controlling the dissipation of solar-wind turbulence.
In balanced Alfv\'enic turbulence, the cascade is highly anisotropic with respect to the background magnetic field and low frequency \citep{matthaeusEvidencePresenceQuasitwodimensional1990,goldreichTheoryInterstellarTurbulence1995,horburyAnisotropicScalingMagnetohydrodynamic2008,chenRecentProgressAstrophysical2016a}.
Standard theories for turbulent dissipation in balanced turbulence therefore favor the perpendicular heating of ions by non-resonant stochastic heating \citep{chandranPerpendicularIonHeating2010b,chandranAlfvenwaveTurbulencePerpendicular2010b}, rather than by cyclotron-resonant heating \citep{isenbergPreferentialAccelerationHeating1983a,hollwegCyclotronResonanceCoronal1999,cranmerCoronalHolesHighSpeed2002}.
In imbalanced turbulence, however, the ``helicity barrier'' effect \citep{meyrandViolationZerothLaw2021} throttles the cascade energy flux, allowing only the subdominant balanced portion of the cascade to reach perpendicular scales $k^{-1}_\perp$ that are smaller than the proton Larmor radius $\rho_{\rm p}$. As a result, the dominantly imbalanced fluctuations in the inertial range grow in amplitude. This in turn decreases the nonlinear turnover time, resulting in smaller parallel scales and therefore higher-frequency fluctuations. 
Hybrid-kinetic simulations show that this evolution promotes cyclotron-resonant heating of protons \citep{squireHighfrequencyHeatingSolar2022} and, even more strongly, minor ions \citep{zhangExtremeHeatingMinor2025}.

Vital to our phenomenology is the distinction between heating by the oblique fluctuations that comprise the majority of the turbulent cascade and heating facilitated by coherent, circularly polarized, parallel-propagating PCWs.
The latter have been conjectured to arise through a process known as ``quasilinear focusing'' \citep{chandranResonantInteractionsProtons2010,isenbergKineticModelSolar2011a}. 
During this process, protons, which compose the bulk of the plasma, are anisotropically heated via cyclotron resonance when there are high-frequency, oblique fluctuations in the cascade. 
The resulting non-Maxwellian proton VDF then becomes unstable to the emission of parallel PCWs, leading to a transfer of energy from oblique cyclotron-frequency fluctuations to parallel PCWs. Hybrid-kinetic simulations of imbalanced turbulence have self-consistently demonstrated quasilinear focusing; in this case, the helicity barrier leads to an increase in the frequencies of oblique, inertial-range fluctuations, which then perpendicularly heat the proton VDF to such an extent that it becomes unstable to the emission of parallel PCWs \citep{squireHighfrequencyHeatingSolar2022,zhangExtremeHeatingMinor2025}.
Consistent with predictions for turbulence governed by the helicity barrier, parallel PCWs in the solar wind are commonly observed to correlate with cross helicity \citep{bowenMediationCollisionlessTurbulent2024,panchalEvidenceLinkTurbulence2025}.
When these high-frequency parallel PCWs control the heating of minor ions, our theory predicts preferential heating of ion species having smaller mass-to-charge ratios, \(A_i/Z_i\), or equivalently larger gyrofrequencies.
Thus, quasilinear focusing in imbalanced turbulence provides a self-consistent mechanism for the inverted scalings of minor-ion thermal speeds with mass-to-charge ratio that are observed in the solar wind, including those shown in Fig.~\ref{fig:tracy2016}.

The remainder of this article is organized as follows.
In \S\ref{sec:theory}, we present the quasilinear theory underlying our phenomenology, first reviewing quasilinear diffusion and focusing before deriving a predicted scaling law for how minor-ion heating rates should depend on ion mass and charge in \S\ref{sec:ql-minor-heat}.
In \S\ref{sec:numerical-model}, we introduce the numerical model that we use to solve the hybrid-kinetic equations to drive balanced and imbalanced turbulence.
We also describe the six simulations analyzed in this work: four new simulations with initial proton plasma beta parameters \(\beta_{\mathrm{p}0}=1\) and \(\beta_{\mathrm{p}0}=1/16\), each with six minor-ion species, together with two prior simulations at \(\beta_{\mathrm{p}0}=0.3\) from \citet{zhangExtremeHeatingMinor2025}.
In \S\ref{sec:time-evol-turb}, we describe the time evolution of the turbulence and ion kinetics, including the development of the helicity barrier, the growth of parallel PCWs through quasilinear focusing, and the resulting minor-ion temperatures and temperature anisotropies.
In \S\ref{sec:mass-charge-scalings}, we show how parallel PCWs can invert the dependence of minor-ion heating rates on mass-to-charge ratio, reproducing the negative correlations sometimes seen in solar-wind and coronal-hole data.
We then use the agreement between the simulations and quasilinear theory to prescribe how proton and minor-ion heating by Alfv\'enic turbulence depend on \(\beta_{\mathrm{p}0}\), the normalized cross helicity \(\sigma_{\mathrm{c}}\) that quantifies the turbulence imbalance, and the masses and charges of different ion species.

\section{Theory}
\label{sec:theory}

In this section, we use quasilinear theory to develop a predictive model for how minor-ion heating depends on the ion-to-proton mass ratio, \(A_i\), and charge ratio, \(Z_i\), defined in Eq.~\eqref{eq:aizi}.
Quasilinear theory describes how a spectrum of linear waves heat each plasma species through resonant wave--particle interactions \citep{isenbergPreferentialAccelerationHeating1983a,isenbergResonantAccelerationHeating1984}. The application to turbulence is discussed in \citet{johnstonQuasilinearTheoryPerpendicular2025}, where they provide evidence that quasilinear theory provides an accurate description of ion heating in imbalanced turbulence, because nonlinear broadening of the frequency spectrum by wave--wave interactions decreases with increasing imbalance.

The slow evolution of a spatially averaged, gyrotropic ion VDF, \(f_{i 0}(t,w_{\parallel },w_{\perp }  ) \), due to fluctuations obeying a linear dispersion relation, \(\omega_{\bs{k}}=\omega(\bb{k})\), is given by \citep{kennelVelocitySpaceDiffusion1966}
\begin{align}
\label{eq:quasilinear-diffusion}
\pdv{f_{i0}}{t} &= \lim_{\mathcal{V}\to\infty} \sum_{n=-\infty}^{\infty} \frac{\pi q_i^2}{m_i^2} \int \frac{\rmd{\bb{k}}}{\mathcal{V}} \nonumber \\
&\quad \times \frac{1}{w_{\perp}}\mathcal{G} \biggl[ w_{\perp} \delta(\omega_{\bs{k}}-k_{\parallel}w_{\parallel}-n\Omega_i) \abs{\psi_{n,\bs{k}}}^2 \mathcal{G} f_{i0} \biggr],
\end{align}
where \(w_{\parallel } \) and \(w_{\perp } \) are peculiar velocities in the plasma frame in the directions
parallel and perpendicular to the local magnetic field; \(\mathcal{V}\) is an integration volume; \(\Omega_{i} = \Omega_{\rm p} Z_i / A_i \) is the gyrofrequency of ion species \(i\); and \(\omega _{\bs{k}} = k_{\parallel } v_{\mathrm{ph},\bs{k}}   \) is the real frequency of
a mode with wave vector \(\bb{k}\) and phase speed \(v_{\mathrm{ph},\bs{k}} \).
\(\psi_{n,\bs{k}}  \) is a weighting function of general wave polarizations, and is given by
\begin{align}
\label{eq:polarization}
\psi_{n,\bs{k}}=
&\frac{1}{\sqrt{2}} [E_{\bs{k}}^{-} \rme^{\imag \phi} \rmJ_{n+1}(\lambda_i)+
E_{\bs{k}}^{+} \rme^{- \imag \phi} \rmJ_{n-1}(\lambda_i)] \nonumber\\
&+\frac{w_{\parallel}}{w_{\perp}}E_{\parallel,\bs{k}}\rmJ_{n}(\lambda_i),
\end{align}
where \(\lambda_i\doteq k_{\perp}w_{\perp}/\Omega_i\), \(\phi\) is the azimuthal angle in \(\bb{k}\) space, \(\rmJ_n\) is the Bessel function of the first kind, and \(\bb{E}_{\bs{k}}\) is the Fourier-transformed electric field with circularly polarized (perpendicular) components given by \(E_{\bs{k}}^{\pm}\doteq  (E_{y,\bs{k}}\pm \imag E_{z,\bs{k}}) / \sqrt{2}\) \citep{chandranResonantInteractionsProtons2010,kennelVelocitySpaceDiffusion1966}.
Equation~\eqref{eq:quasilinear-diffusion} is a diffusion equation in velocity space,
with the two applications of the operator
\begin{equation}
\label{eq:quasilinear-operator}
 \mathcal{G} \doteq \biggl(1- \frac{w_{\parallel }}{v_{\mathrm{ph},\bs{k}}  } \biggr) \pdv{}{w_{ \perp } } +\frac{w_{\perp } }{ v_{\mathrm{ph},\bs{k}} } \pdv{}{w_{\parallel } }
\end{equation}
leading to drag and diffusion in velocity space.

Only resonant particles, whose parallel velocities satisfy the resonance condition
\begin{equation}
\label{eq:resonance-condition}
\omega _{\bs{k}} -k_{\parallel } w_{\parallel } -n \Omega _{i} = 0,
\end{equation}
interact with the waves.
Diffusion acts to minimize velocity-space gradients until
\(\mathcal{G}[f_{i0} ]=0\), flattening the VDF along resonant
contours in velocity space whose shapes are defined by the level sets of any function \(\Xi(w_{\perp },w_{\parallel }  )\) along
which \(\mathcal{G}[\Xi]=0\).
One determines \(\Xi(w_{\perp },w_{\parallel }  )\) by integrating \(\mathcal{G}=0\)
in \(w_{\perp } \) and \(w_{\parallel } \) and using the resonance condition \eqref{eq:resonance-condition} to relate \(k_{\parallel } \) to \(w_{\parallel } \) (or
vice versa). For the cyclotron resonance \(n=1\), this provides the relation \citep{rowlandsQuasilinearTheoryPlasma1966,gendrinPitchAngleDiffusion1968,isenbergDispersiveAnalysisBispherical1996}
\begin{equation}
\label{eq:scattering-contours}
w_{\parallel }^2 + w_{\perp }^2 - 2 \int \rmd{w_{\parallel } }\, v_{\mathrm{ph},\bs{k}}(k_{\parallel } (w_{\parallel }) ) = \const .    \end{equation}
These level sets define \emph{scattering contours} that represent conservation of energy in the wave frame at any given \(k_{\parallel } \).
Viewed in velocity space, the instantaneous radius of curvature for a level set
at a given \(w_{\parallel } (k_{\parallel } ) \) depends on the phase speed of the resonant wave.
Resonance with faster waves increases this radius, and therefore decreases the
curvature in velocity space, leading to scattering contours that are steeper in \(w_{\perp } \).

\begin{figure}[htbp]
    \centering
    \includegraphics[width=0.95\columnwidth]{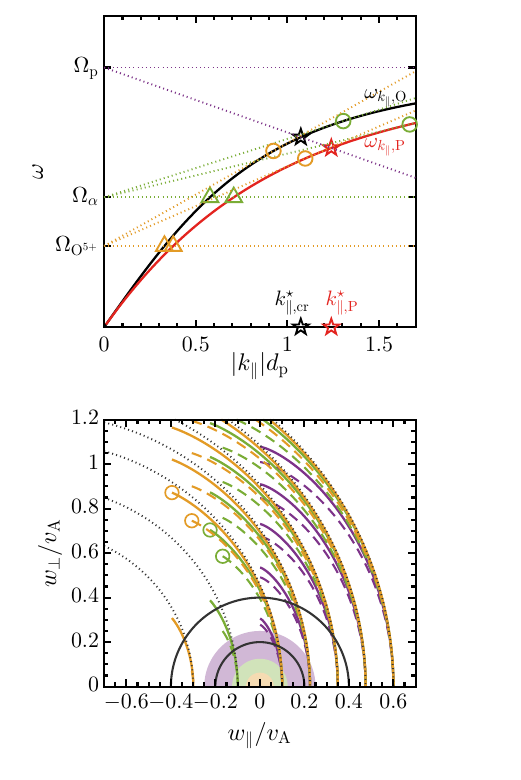}
    \caption{Top panel: Dispersion relations for oblique PCWs, \(\omega_{k_{\parallel},{\rm O}}\) (black), and parallel PCWs, \(\omega_{k_{\parallel},{\rm P}}\) (red), plotted as functions of \(|k_{\parallel}|\), with \(k_{\parallel}<0\) in our convention.
Dotted lines show \(\Omega_i-|k_{\parallel}|w_{\parallel}\) for protons (purple), alphas (green), and \({\rm O}^{5+}\) (orange); their intersections with \(\omega_{k_{\parallel}}\) give the resonant wavenumbers \(k_{\parallel,{\rm res}}\) satisfying Eq.~\eqref{eq:resonance-condition} for oblique or parallel waves.
Horizontal dotted lines correspond to \(w_{\parallel}=0\); the diagonal proton line corresponds to \(w_{\parallel}=v_{{\rm th,p}0}\) for \(\beta_{{\rm p}0}=1/16\), giving \(v_{{\rm th,p}0}= 0.25 v_{\rm A}\); and the diagonal alpha and \({\rm O}^{5+}\) lines indicate \(w_{\parallel,\min}<0\) for resonance with oblique or parallel PCWs. \(k_{\parallel,{\rm cr}}^{\star}\) and \(k_{\parallel,{\rm P}}^{\star}\) denote the intersections of the diagonal proton line with \(\omega_{k_{\parallel,{\mathrm O}}}\) and \(\omega_{k_{\parallel,{\mathrm P}}}\), respectively, and are described in \S\ref{sec:ql-focusing}.
Bottom panel: Scattering contours in velocity space for protons (purple), alphas (green), and \({\rm O}^{5+}\) (orange) that are resonant with oblique (solid) or parallel (dashed) PCWs.
For each species, \(w_{\parallel,\min}\) for resonance with either wave mode is marked by circles at the ends of the contours that start from \(w_{\parallel}=+v_{\rm A}\); the corresponding \(k_{\parallel,{\rm res}}(w_{\parallel,\min})\) values are marked by circles in the top panel.
Non-dispersive \(\omega=k_{\parallel}v_{\rm A}\) contours (dotted) and constant-energy contours (black, solid) are semicircles centered on the \(w_{\parallel}\)-axis at \(-v_{\rm A}\) and \(0\), respectively.
Colored shaded regions are semicircles of radius \(v_{{\rm th},i0}=\sqrt{\beta_{{\rm p}0}/m_i}\) at \(\beta_{{\rm p}0}=1/16\), proportional to the extent of the initial thermal core of each species.}
    \label{fig:res-cont}
\end{figure}

To compute scattering contours that describe cyclotron-resonant ion heating by oblique Alfv\'en/PCWs,
we adopt the cold plasma dispersion relation for \({k_{\perp } \gg k_{\parallel }}  \)
\citep{stixWavesPlasmas1992,isenbergSelfconsistentMarginallyStable2012}, 
\begin{equation}
\label{eq:oblique-pcw}
\omega_{k_{\parallel },\mathrm{O} }  =\Omega _{\mathrm{p}}  \frac{k_{\parallel }d_{\mathrm{p}} }{\sqrt{1+(k_{\parallel } d_{\mathrm{p}}) ^2}},
\end{equation}
where the subscript ``\(\mathrm{O}\)'' denotes oblique waves, and \(d_{\rm p}\) is the proton inertial length. A cold-plasma dispersion relation should be a reasonable approximation for sufficiently low proton beta.
In the limit \(k_{\parallel}d_{\mathrm{p}}\ll1\), Eq.~\eqref{eq:oblique-pcw} reduces to \(\omega_{k_{\parallel},\mathrm{O}}\simeq k_{\parallel}v_{\mathrm{A}}\), describing non-dispersive, oblique Alfv\'en waves, whose linear physics accurately describes most of the inertial-range fluctuations in a strong, Alfv\'enic cascade.
We also consider parallel PCWs, \(k_{\perp}=0\), whose cold-plasma dispersion relation is \citep{stixWavesPlasmas1992,hollwegGenerationFastSolar2002}
\begin{equation}
\label{eq:prl-pcw}
\omega_{k_{\parallel }, \mathrm{P} }  = \frac{\Omega_{\mathrm{p}}}{2}   (k_{\parallel } d_{\mathrm{p}})^2  \biggl( \sqrt{1+4 (k_{\parallel } d_{\mathrm{p}})^{-2} } -1\biggr),
\end{equation}
where the subscript ``\(\mathrm{P}\)'' denotes parallel waves.
The top panel of Fig.~\ref{fig:res-cont} shows the oblique and parallel PCW dispersion relations, \(\omega_{k_{\parallel},\mathrm{O}}\) and \(\omega_{k_{\parallel},\mathrm{P}}\), as black and red curves, respectively.
Their intersections with the dotted lines, \(\Omega_i-|k_{\parallel}|w_{\parallel}\), give the resonant wavenumbers \(k_{\parallel,\mathrm{res}}(w_{\parallel})\) satisfying Eq.~\eqref{eq:resonance-condition} for $n=1$ and different ion species.
Writing the dimensionless parallel wavenumber as \(\bar{k}_{\parallel} \doteq k_{\parallel} d_{\rm p} \), Eqs.~\eqref{eq:resonance-condition}--\eqref{eq:prl-pcw} give the scattering contours
\begin{gather}
\label{eq:scattering-contours-form}
\frac{w_{\perp }^2}{v_{\rm A}^2} + \frac{Z_i}{A_i}\bar{k}_{\parallel }^{-2}\biggl\{\frac{Z_i}{A_i} + \frac{2 \bar{k}_{\parallel }^{3}}{\sqrt{1+\bar{k}_{\parallel }^2}}   \biggr\}      = \const      \\
\frac{w_{\perp }^2}{v_{\rm A}^2} + \frac{Z_i}{A_i} \biggl\{\frac{Z_i}{A_i} \bar{k}_{\parallel }^{-2} - \ln\Bigl[1+  \sqrt{1+4 \bar{k}_{\parallel }^{-2} }    \Bigr] \biggr\}  = \const,
\end{gather}
for oblique and parallel PCWs, respectively \citep[e.g.;][]{isenbergDispersiveAnalysisBispherical1996,isenbergKineticShellModel2001,isenbergKineticShellModel2004,isenbergKineticModelSolar2011a}.

In the bottom panel of Fig.~\ref{fig:res-cont}, we plot scattering contours for protons (purple), alphas (green), and \({\rm O}^{5+}\) (orange) resonant with oblique (solid) and parallel (dashed) PCWs.
Quasilinear diffusion flattens ion VDFs along these contours.
When the VDF decreases along a scattering contour in the direction of increasing kinetic energy, diffusion carries particles across constant-energy shells (black solid semicircles) toward larger \(w^2=w_{\parallel}^2+w_{\perp}^2\), thereby heating the resonant ions.
The non-dispersive Alfv\'en-wave limit, \(\omega=k_{\parallel}v_{\rm A}\) for \(k_{\parallel}d_{\rm p}\ll1\), is shown by the dotted contours.
This limit becomes a poorer approximation at larger resonant \(|k_{\parallel}|\), where the waves are more dispersive.
For a fixed resonant \(w_{\parallel}\), ions with smaller mass-to-charge ratio, \(A_i/Z_i\), resonate with waves at larger \(|k_{\parallel}|\); for any given species, the resonant \(|k_{\parallel}|\) also increases as \(w_{\parallel}\) becomes more negative.
At these larger \(|k_{\parallel}|\), dispersion reduces the phase speeds of both oblique and parallel PCWs, making their scattering contours shallower in \(w_{\perp}\) than the non-dispersive contours.
\citet{yergerCyclotronBreakingMechanism2026} prove that, for a given resonant \(w_{\parallel}\), parallel PCWs have lower phase speeds than the corresponding oblique PCWs, \(\omega_{k_{\parallel,{\rm res},{\rm O}},{\rm O}}/k_{\parallel,{\rm res,O}}>\omega_{k_{\parallel,{\rm res},{\rm P}},{\rm P}}/k_{\parallel,{\rm res,P}}\). Parallel PCWs therefore possess even shallower contours than oblique PCWs, as per the dashed lines in Fig.~\ref{fig:res-cont}.
Likewise, the contours become progressively shallower from \({\rm O}^{5+}\) to alphas to protons, reflecting the smaller \(A_i/Z_i\) and larger resonant \(|k_{\parallel}|\) of the lighter, more rapidly gyrating ions.

For Alfv\'enic fluctuations propagating antiparallel to the local magnetic field, \(k_{\parallel}<0\) (as for the Elsasser field \(\bb{z}^{+}\) in our convention; see Eq.~\eqref{eq:elsasser}), there also exists a minimum parallel velocity, \(w_{\parallel,\min}\leq0\), below which no cyclotron (\(n=1\)) resonance occurs.
For protons, \(w_{\parallel,\min}=0\), because the maximum PCW frequency is the proton gyrofrequency, \(\Omega_{\rm p}\).
For minor ions, \(\Omega_i<\Omega_{\rm p}\), so \(w_{\parallel,\min}\) is set by the steepest positive-slope dotted line in the top panel of Fig.~\ref{fig:res-cont} that still intersects \(\omega_{k_{\parallel}}\), which occurs where the line is tangent to the dispersion relation, as indicated by the circles.
Analytically, this condition is \(v_{{\rm g},\bs{k}}=w_{\parallel,\min}(k_{\parallel,\rm res})\), where \(v_{{\rm g},\bs{k}}=\partial\omega_{\bs{k}}/\partial k_{\parallel}\) is the wave's parallel group speed.
Thus, \(w_{\parallel,\min}\) marks the endpoint of the scattering contours, shown by colored circles at the ends of the alpha and \({\rm O}^{5+}\) contours in the bottom panel of Fig.~\ref{fig:res-cont}.
Ions with larger mass-to-charge ratio, \(A_i/Z_i\), have smaller gyrofrequency and larger \(|w_{\parallel,\min}|\), so their contours extend farther into \(w_{\parallel}<0\), as seen by comparing the \({\rm O}^{5+}\) contours (orange) with the alpha contours (green).
Heating therefore occurs over a broader range of \(w_{\parallel}\) for ions with larger \(A_i/Z_i\).

\subsection{Quasilinear focusing}
\label{sec:ql-focusing}

Efficient cyclotron-resonant perpendicular heating of bulk protons by Alfv\'enic turbulence can drive an instability that re-emits the damped energy as parallel PCWs, a process termed \emph{quasilinear focusing} \citep{chandranResonantInteractionsProtons2010,isenbergKineticModelSolar2011a}.
When resonant processes dominate, quasilinear diffusion \eqref{eq:quasilinear-diffusion} flattens initially Maxwellian ion VDFs along the resonant scattering contours of the oblique Alfv\'{e}n/PCW fluctuations.
The resulting proton VDF develops velocity-space gradients along the shallower parallel-PCW contours (bottom panel of Fig.~\ref{fig:res-cont}), causing diffusion toward lower kinetic energy.
This process damps oblique PCW energy and re-emits, or ``focuses'', it into parallel PCWs.

Strong Alfv\'enic turbulence is expected to satisfy scale-by-scale critical balance between the linear propagation time and the nonlinear cascade time, such that \(k_{\parallel}v_{\rm A}\sim k_{\perp}\delta u_{\perp}\) \citep{goldreichTheoryInterstellarTurbulence1995,boldyrevSpectrumMagnetohydrodynamicTurbulence2006,malletRefinedCriticalBalance2015}.
As fluctuation amplitudes decrease toward smaller scales, critical balance implies that fluctuations become increasingly anisotropic, \(k_{\parallel}/k_{\perp}\sim\delta u_{\perp}/v_{\rm A}\).
Thus, the fluctuation power available at the large \(k_{\parallel}\) required by the resonance condition \eqref{eq:resonance-condition} to resonantly heat the bulk of the proton VDF decreases rapidly.
In energetically balanced turbulence, nonlinear interactions between counter-propagating fluctuations further produce a broad temporal frequency spectrum, rather than a narrow spectrum concentrated near a wave dispersion relation \citep{johnstonQuasilinearTheoryPerpendicular2025}.
Thus, resonant proton-cyclotron heating is weak in balanced Alfv\'enic turbulence, and quasilinear focusing through cyclotron resonance is not expected.

Because energetically imbalanced turbulence has a narrower frequency spectrum, resonant proton-cyclotron heating is enhanced when the cross helicity \(\sigma_{\rm c}\ne0\).
Additionally, in imbalanced turbulence, the ``helicity barrier'' disallows a constant-flux cascade through to sub-proton-Larmor scales, \(k_{\perp}\rho_{\rm p}\gtrsim1\), instead allowing only the balanced portion of the energy flux to reach smaller, kinetic scales \citep{meyrandViolationZerothLaw2021}.
This ``barrier'' produces a steep transition range in the \(k_{\perp}\) spectrum, where the fluctuation energy rapidly decreases until only the smaller balanced component remains.
Consistent with this prediction, solar-wind observations show that steeper transition-range spectra are associated with larger cross helicity \citep{mcintyreEvidenceHelicityBarrier2025}.
We denote the start of this transition range by \(k_{\perp}^{\star}\), whose scaling with imbalance is predicted by the helicity barrier to satisfy \(k_{\perp}^{\star}\rho_{\rm p}\sim(1-\sigma_{\rm c})^{1/4}\) \citep{meyrandViolationZerothLaw2021,squireElectronIonHeatingPartition2023,adkinsTurbulentHeatingCollisionless2025}.

Suppose initially that fluctuation amplitudes are weak enough that, before the transition range, the largest \(k_{\parallel}\) with non-negligible power resonates only with protons far in the tail of the VDF, \(|w_{\parallel}|\gg v_{\rm th,p}\).
Then cyclotron-resonant proton heating is initially insufficient to dissipate the cascade flux, while the helicity barrier prevents most of that flux from accessing electron-heating channels at yet smaller scales.
The dominantly imbalanced inertial-range fluctuations therefore grow in amplitude, increasing their parallel wavenumber through critical balance, \(k_{\parallel}/k_{\perp}\sim\delta u_{\perp}/v_{\rm A}\sim\delta B_\perp/B_0\).
Once sufficient fluctuation power reaches parallel scales resonant with the thermal bulk of the proton VDF, the fluctuations saturate through proton-cyclotron-resonant heating, flattening the proton VDF along oblique contours up to \(w_{\parallel}\approx0\).
We define the critical parallel wavenumber for this saturation as the oblique-PCW resonant wavenumber for protons with \(w_{\parallel}=v_{\rm th,p}\),
\(k_{\parallel,{\rm cr}}^{\star}\doteq k_{\parallel,{\rm res},{\rm O}}(w_{\parallel}=v_{\rm th,p})\), marked for \(\beta_{{\rm p}0}=1/16\) in the top panel of Fig.~\ref{fig:res-cont}.
This dissipation of oblique fluctuations enables strong quasilinear focusing, re-emitting energy into parallel PCWs concentrated near a coresonant scale, which we define as \(k_{\parallel,{\rm P}}^{\star}\) and also mark in Fig.~\ref{fig:res-cont}.

Quasilinear diffusion by a given wave population weakens as the ion VDF flattens along that population's scattering contours; in the limit \(\mathcal{G}f_{i0}=0\), the VDF is flat along those contours and that wave population no longer heats the ions.
Consider a near-saturated state in which the energetically dominant oblique PCWs have already flattened the minor-ion VDFs along oblique-PCW contours, so that \(\mathcal{G}_{\rm O}f_{i0}\) is small.
Here, \(\mathcal{G}_{\rm O}\) and \(\mathcal{G}_{\rm P}\) denote the operator \(\mathcal{G}\) evaluated along oblique- and parallel-PCW scattering contours, respectively.
Because parallel-PCW contours are generally shallower than oblique-PCW contours in a cold plasma, as shown in Fig.~\ref{fig:res-cont}, the same VDF can retain much larger gradients along the parallel-PCW contours, \(|\mathcal{G}_{\rm P}f_{i0}|\gg|\mathcal{G}_{\rm O}f_{i0}|\).
Diffusion along these shallower contours proceeds toward lower kinetic energy and emits parallel PCWs, but it also moves particles across otherwise nearly saturated oblique-PCW contours.
The energetically dominant oblique cascade can then rapidly flatten the VDF again along oblique contours at higher kinetic energy, producing a net enhancement of minor-ion heating.
This cross-contour diffusive process is analogous to that described by \citet{chandranResonantInteractionsProtons2010} for protons \citep[see also][]{isenbergKineticModelSolar2011a}.
When quasilinear focusing generates parallel PCWs of sufficient amplitude, this cross-contour process becomes the bottleneck for enhanced heating, shifting the species-dependent heating from being controlled by the oblique-PCW spectrum to being controlled by the parallel-PCW spectrum.

\subsection{A scaling theory for quasilinear heating of minor ions}
\label{sec:ql-minor-heat}

To determine how minor-ion heating is predicted to depend on mass and charge, we take the second velocity moment of Eq.~\eqref{eq:quasilinear-diffusion}.
We evaluate the integral over \(k_{\parallel}\) using \(\delta[h(x)]=\sum_j \delta(x-x_j)/|h'(x_j)|\), where \(x_j\) are the roots of \(h(x)\), and then integrate by parts in velocity.
This gives the mass-normalized quasilinear heating rate
\begin{align}
\label{eq:ql-heating-rate}
\frac{Q_{i}}{m_{i}} &\doteq \int \rmd{\bb{r}}\, \int \rmd{\bb{w}}\, \frac{1}{2} w^2 \pdv{f_{i0}}{t} \nonumber \\
\mbox{} &= - \frac{\pi^2 q_i^2}{m_i^2} \sum_{n=-\infty }^{\infty }   \int_{-\infty  }^{\infty } \rmd{w_{\parallel } }\, \int \rmd{\bb{w}_{\perp } }\, \nonumber \\
&\quad \quad \quad \quad \frac{w_{\perp }}{{|v_{{\rm g},k_{\parallel ,\mathrm{res}} } - w_{\parallel }  |}} \mathcal{I}_{n} (k_{\parallel , \mathrm{res}} ) \mathcal{G}_{k_{\parallel , \mathrm{res}} } f_{i0},
\end{align}
where \({\int \rmd{\bb{w}_{\perp}}\{\ldots\}=\int_0^{\infty}w_{\perp}\rmd{w_{\perp}}\{\ldots\}}\), \(k_{\parallel,\mathrm{res}}(w_{\parallel})\) is determined by the resonance condition \eqref{eq:resonance-condition}, and
\({\mathcal{I}_{n} (k_{\parallel }) \doteq   \int \rmd{\bb{k}_{\perp } }\, \abs{\psi _{n,\bs{k}} }^2 } \).
In writing \eqref{eq:ql-heating-rate}, we have neglected the \(k_{\perp}\) dependence of the dispersion relation, taking \(\omega_{\bs{k}}=\omega(k_{\parallel})\), and the group speed to be \(v_{{\rm g},k_{\parallel}} = \partial \omega_{k_{\parallel}} / \partial k_{\parallel}\).

From Eq.~\eqref{eq:ql-heating-rate}, the quasilinear heating rate depends on ion mass and charge not only explicitly through the \(q_i^2/m_i^2\) coefficient, but also implicitly through \(k_{\parallel,\mathrm{res}}(\Omega_i,w_{\parallel})\) via Eq.~\eqref{eq:resonance-condition}.
The relative contribution of a given \(k_{\parallel,\mathrm{res}}(\Omega_i,w_{\parallel})\) to the integral over \(w_{\parallel}\) depends on both the electric-field power, \(\mathcal{I}_n[k_{\parallel,\mathrm{res}}(w_{\parallel})]\), and the degree to which the ion VDF has been flattened along the relevant scattering contour, \(\mathcal{G}_{k_{\parallel,\mathrm{res}}(w_{\parallel})}f_{i0}(w_{\parallel},w_{\perp})\).
The latter is generally difficult to estimate heuristically.
However, in two regimes relevant to our hybrid-kinetic simulations, we argue that \(\mathcal{G}_{k_{\parallel,\mathrm{res}}}f_{i0}\) does not contribute to the leading-order mass--charge dependence of \(Q_i\).
This allows us to estimate the \(\int\rmd{\bb{w}_{\perp}}\{\ldots\}\) integral in Eq.~\eqref{eq:ql-heating-rate}. These regimes are:

\begin{enumerate}
\item In the first regime, relevant to the initial stages of heating, all minor ions are nearly isothermal with one another and have approximately Maxwellian VDFs.
At low plasma beta, and especially for heavier ions, the scattering contours over the region of velocity space occupied by the slow ion cores are nearly vertical (bottom panel of Fig.~\ref{fig:res-cont}), corresponding to nearly pure perpendicular heating.
Moreover, \(\mathcal{G}\), as given in Eq.~\eqref{eq:quasilinear-operator}, is dominated by \(\partial / \partial w_{\perp}\) when \(|w|\ll v_{{\rm ph},k_{\parallel}}\).
Thus, \(\mathcal{G}_{k_{\parallel,\mathrm{res}}(w_{\parallel})}f_{i0}\sim f_{i0}/v_{{\rm th},\perp i}\), and so
\begin{equation}
\label{eq:regime1}
\int \rmd{\bb{w}_{\perp}}\, w_{\perp}\mathcal{G}_{k_{\parallel,\mathrm{res}}}f_{i0}
\sim f_{i0}(w_{\parallel})
\end{equation}
is independent of both mass and charge.
\item In the second regime, relevant to late times in our numerical simulations, minor-ion VDFs are nearly flattened along resonant scattering contours.
Apart from their extent into \(w_{\parallel}<0\) and modest dispersive effects, these contours are qualitatively similar in velocity space between minor-ion species (Fig.~\ref{fig:res-cont}).
The minor-ion VDFs therefore occupy similar regions of velocity space at late times, largely independent of ion mass.
To leading order, differences in thermal speed are then small, and minor-ion temperatures become approximately mass proportional.
If, in this near-saturated state, gradients of \(f_{i0}\) along the contours are approximately uniform between species, or depend only weakly on \(v_{{\rm th},i}\), then
\begin{equation}
\label{eq:regime2}
\int \rmd{\bb{w}_{\perp}}\, w_{\perp}\mathcal{G}_{k_{\parallel,\mathrm{res}}} f_{i0}
\sim f_{i0}(w_{\parallel})\,v_{{\rm th},\perp i}.
\end{equation}
Because the thermal speeds of the minor-ion species are similar in this regime, this term contributes negligibly to the leading-order mass--charge dependence of \(Q_i\).
\end{enumerate}

To evaluate the \(\int \rmd{w_{\parallel}}\{\ldots\}\) integral, we must specify both \(k_{\parallel,\mathrm{res}}(\Omega_i,w_{\parallel})\) and the power in the relevant wave mode.
In general, \(k_{\parallel,\mathrm{res}}\) must be found by solving the resonance condition \eqref{eq:resonance-condition} numerically for each \(w_{\parallel}\).
For a rough comparison between minor-ion species, we instead evaluate the terms in Eq.~\eqref{eq:ql-heating-rate} at \(w_{\parallel}=0\), where the VDFs are most populous. We maintain this assumption for our phenomenological predictions throughout the remainder of this article.
At this velocity,
\begin{equation}
\label{eq:kprlres}
k_{\parallel,\mathrm{res}}(\Omega_i,w_{\parallel}=0)
=
\frac{\Omega_i}{v_{\mathrm{ph},k_{\parallel,\mathrm{res}}}},
\end{equation}
corresponding to the intersections marked by triangles in the top panel of Fig.~\ref{fig:res-cont}.
This approximation is best for cold ion VDFs with small thermal speeds, \(v_{{\rm th},i}\ll v_{\rm A}\), and therefore applies most directly at low plasma beta and, initially, to heavier minor-ion species.
It does not apply to protons, for which \(w_{\parallel,\min}=0\); as \(w_{\parallel}\to0\), the resonant wavenumber diverges, \(k_{\parallel,\mathrm{res}}\to\infty\), and the available wave power vanishes.

For cyclotron-resonant (\(n=1\)) heating of ions with \(w_{\parallel}=0\) and \(w_{\perp}>0\), the polarization function \(\psi_{1,\bs{k}}\) in Eq.~\eqref{eq:polarization} contains only perpendicular components of the electric field.
We therefore approximate \(\mathcal{I}_1(k_{\parallel})\) as the one-dimensional \(k_{\parallel}\) spectrum of perpendicular electric-field energy in the relevant wave mode, \(\mathcal{E}_{E_{\perp}}(k_{\parallel}) = \mathcal{E}_{E_{y}}(k_{\parallel}) + \mathcal{E}_{E_{z}}(k_{\parallel})\), normalized such that
\(
\int \rmd k_{\parallel}\, \mathcal{E}_{E_{\perp}}(k_{\parallel})
=
\int \rmd k_{\parallel}\rmd k_{\perp}\,
\mathcal{E}_{E_{\perp}}(k_{\parallel},k_{\perp})  
\) is the electric energy density of the mode.
Assuming that this spectrum obeys a power-law scaling over the relevant resonant range, we write
\begin{equation}
  \label{eq:espectrum}
\mathcal{I}_1(k_{\parallel}) \sim \mathcal{E}_{E_{\perp}}(k_{\parallel}) \sim k_{\parallel}^{-\eta},
\end{equation}
where we introduce the spectral exponent \(\eta\).
In general, the spectra associated with parallel and oblique PCWs can have different slopes in \(k_{\parallel}\); when the wave mode is known, we denote these explicitly by \(\eta_{\rm P}\) and \(\eta_{\rm O}\), respectively.

Substituting the wave-power scaling \eqref{eq:espectrum} into Eq.~\eqref{eq:ql-heating-rate}, and using the dependence of \(k_{\parallel,\mathrm{res}}\) on the minor-ion cyclotron frequency, we find that the heating rate should scale with ion mass and charge as
\begin{equation}
\label{eq:ql-heat-rate-disp}
Q_i \propto A_i \left(\frac{A_i}{Z_i}\right)^{\eta-2}
\frac{v_{\mathrm{ph},k_{\parallel,\mathrm{res}}}^{\eta}}{v_{{\rm g},k_{\parallel,\mathrm{res}}}}.
\end{equation}
In the non-dispersive limit, \(v_{\mathrm{ph}}\) and \(v_{\rm g}\) are both equal to \(v_{\rm A}\) and independent of \(k_{\parallel}\), such that
\({Q_i \propto A_i (A_i / Z_i)^{\eta -2} } \).
Thus, a conservative-flux cascade of non-dispersive Alfv\'enic fluctuations, for which \(\eta=2\), predicts only mass-proportional minor-ion heating.
If the electric-field spectrum of the relevant wave mode is steeper (shallower) than \(\eta=2\), the heating rate becomes an increasing (decreasing) function of mass-to-charge ratio.

The relevant wave modes in the two regimes of interest are: in both regimes, oblique Alfv\'en/PCWs that occupy most of the inertial-range cascade; and, in the second regime, coherent parallel-propagating PCWs that allow diffusion across saturated oblique contours when sufficiently energetic.
At \(k_{\parallel,\mathrm{res}}(\Omega_i,w_{\parallel}=0)\), their respective phase speeds, from Eqs.~\eqref{eq:oblique-pcw} and \eqref{eq:prl-pcw}, and group speeds scale with minor-ion parameters as
\begin{align}
\label{eq:opcw-phase-speed}
v_{\mathrm{ph},k_{\parallel,\mathrm{res}},\mathrm{O}} &= v_{\rm A}\left(1-\frac{Z_i^2}{A_i^2}\right)^{1/2}, \\
\label{eq:pcw-phase-speed}
v_{\mathrm{ph},k_{\parallel,\mathrm{res}},\mathrm{P}} &= v_{\rm A}\left(1-\frac{Z_i}{A_i}\right)^{1/2}, \\
\label{eq:opcw-group-speed}
v_{\mathrm{g},k_{\parallel,\mathrm{res}},\mathrm{O}} &= v_{\rm A}\left(1-\frac{Z_i^2}{A_i^2}\right)^{3/2}, \\
\label{eq:pcw-group-speed}
v_{\mathrm{g},k_{\parallel,\mathrm{res}},\mathrm{P}} &= v_{\rm A}\left(1-\frac{Z_i}{A_i}\right)^{3/2}\left(1-\frac{1}{2}\frac{Z_i}{A_i}\right)^{-1}.
\end{align}
Substituting these expressions into Eq.~\eqref{eq:ql-heat-rate-disp} gives the dispersive heating-rate scalings for each mode
\begin{align}
\label{eq:ql-heat-rate-fullO}
Q_{i,\mathrm{O}} &\propto A_i\left(\frac{A_i}{Z_i}\right)^{\eta_{\rm O}-2}
\left(1-\frac{Z_i^2}{A_i^2}\right)^{(\eta_{\rm O}-3)/2}, \\
\label{eq:ql-heat-rate-full}
Q_{i,\mathrm{P}} &\propto A_i\left(\frac{A_i}{Z_i}\right)^{\eta_{\rm P}-2}
\left(1-\frac{Z_i}{A_i}\right)^{(\eta_{\rm P}-3)/2}
\left(1-\frac{1}{2}\frac{Z_i}{A_i}\right).
\end{align}
Ions with smaller gyrofrequencies resonate with fluctuations at smaller \(k_{\parallel}\), where both oblique and parallel PCWs are less dispersive and \(v_{\mathrm{ph}}\) is closer to \(v_{\rm A}\).
The departure of Eqs.~\eqref{eq:ql-heat-rate-fullO} and \eqref{eq:ql-heat-rate-full} from the non-dispersive scaling \(Q_i\propto A_i(A_i/Z_i)^{\eta-2}\) is therefore smaller for ions with larger mass-to-charge ratio.
Similarly, because parallel PCWs have slower phase speeds than oblique PCWs at a given \(k_{\parallel}\), dispersive corrections are larger for parallel PCWs.

To use this theory in practice, we first identify the wave population that contributes most to the heating and measure its electric-field spectrum over the range of \(k_{\parallel}\) resonant with minor ions at \(w_{\parallel}=0\).
The local spectral slope, \(\mathcal{E}_{E}\sim k_{\parallel}^{-\eta}\), then predicts the leading-order mass--charge exponent, \(a\simeq\eta-2\), with dispersive corrections given by Eqs.~\eqref{eq:ql-heat-rate-fullO} and \eqref{eq:ql-heat-rate-full}.
For imbalanced turbulence, the relevant wave population follows from the stage of the evolution.
At early times, before quasilinear focusing is strong, the resonant fluctuation energy resides primarily in oblique Alfv\'en/PCW fluctuations in the turbulent cascade.
These fluctuations heat the initially Maxwellian ion cores, with a mass--charge dependence given by Eq.~\eqref{eq:ql-heat-rate-fullO}.
At late times, after the ion VDFs have become nearly saturated along oblique scattering contours, quasilinear focusing generates non-negligible energy in coherent parallel PCWs.
These parallel PCWs, concentrated near \(k_{\parallel,{\rm P}}^{\star}\), facilitate further heating through cross-contour diffusion, leading to the mass--charge dependence in Eq.~\eqref{eq:ql-heat-rate-full}.
Differences in the \(k_{\parallel}\) electric-field spectra, \(\mathcal{E}_{E_{\perp}}(k_{\parallel})\), of oblique and parallel PCWs therefore produce different power-law scalings of heating rate with mass-to-charge ratio.
For parallel PCWs, this scaling can invert when their energy peaks above the resonant wavenumbers of some minor ions, \(k_{\parallel,{\rm P}}^{\star}>k_{\parallel,{\rm res}}\), producing a sufficiently shallow effective spectrum, \({\eta<2}\), over the resonant range.

\section{Simulation}
\label{sec:numerical-model}

We test our theoretical predictions for ion heating using a suite of numerical simulations of driven Alfv\'enic turbulence similar to that found in the solar wind.
These simulations follow a local, comoving patch of solar wind in a triply periodic domain that is elongated along a mean magnetic field, \(\bb{B}_0\).
The initial state is homogeneous, with background proton density \(n_{{\rm p}0}\), and consists of bulk protons (\(i={\rm p}\)), electrons, and trace-abundance minor-ion species in thermal equilibrium, \(T_{{\rm p}0}=T_{\rm e0}=T_{i0}\).
This initial temperature equilibrium is motivated by a parcel of plasma originating deep in the corona, where the ion species are assumed to collisionally decouple at \(t=0\), as the plasma is also subjected to forcing by fluctuations near the box scale.
For simplicity, we neglect solar-wind expansion and acceleration, assuming that the corresponding background-evolution timescales are very long compared to the characteristic linear and nonlinear timescales of the turbulence. 
These choices make the simulations controlled local experiments rather than self-consistent models for an expanding plasma in a solar-wind flux tube.
By varying \(\beta_{\mathrm{p}0}\) and \(\sigma_{\rm c}\), which change throughout the solar wind and with heliocentric distance, we use these local simulations to probe ion heating under conditions representative of different solar-wind regions.

Our model equations and numerical approach are identical to those  described in \citet{zhangExtremeHeatingMinor2025}.
Namely, we adopt a hybrid-kinetic approach \citep{byersHybridSimulationsQuasineutral1978,hewettMultidimensionalQuasineutralPlasma1978}, which affords a sufficiently large separation of scales while retaining finite-Larmor-radius effects and allowing for high-frequency fluctuations. The system of multi-ion hybrid-kinetic equations
is given by
\begin{align}
  \label{eqn:hybrid-vlasov}
\pdv{f_i}{t} + \bb{v}\bcdot\grad f_i + \frac{q_i}{m_i} \biggl(\bb{E} + \frac{\bb{v}}{c}\btimes\bb{B} \biggr)\bcdot\pdv{f_i}{\bb{v}} \nonumber\\*
\mbox{} = -\frac{q_i}{m_i} \bb{E}^{U}_{\rm ext}\bcdot \pdv{f_{i} }{\bb{v}} , 
\end{align}
\begin{equation}
  \label{eqn:faraday}
\frac{1}{c} \pdv{\bb{B}}{t}  + \grad\btimes\bb{E} = -\grad\btimes\bb{E}^{B}_{\rm ext}  -\eta_4 \nabla^4\bb{B} ,
\end{equation}
\begin{align}
  \label{eqn:ohm}
\bb{E} = - \frac{\sum_i \Lambda_i q_i n_i \bb{u}_i}{\sum_i \Lambda_i q_i n_i c }\btimes\bb{B}+ &\frac{(\grad\btimes\bb{B}) \btimes\bb{B}}{4\pi\sum_i \Lambda_i q_i n_i}  \\
&-\frac{T_{\rm e}}{e}\frac{\grad\sum_i \Lambda_i q_i n_i}{\sum_i \Lambda_i q_i n_i} \nonumber ,
\end{align}
where \(q_i = Z_i e\) and \(m_i = A_i m_{\mathrm p}\) are the charge and mass of ion species
\(i\) expressed in terms of the proton charge ($e$) and proton mass ($m_{\rm p}$), \(c\) is the speed of light, \(\bb{E}\) is the electric field, and 
\(\bb{B}\) is the magnetic field. The parameter \(\Lambda_i\) equals unity for \emph{active} ion species and zero for \emph{passive} ion species. In all simulations presented, only protons are active. The external electric fields 
\(\bb{E}_{\mathrm{ext}}^{B}  \) and \(\bb{E}_{\mathrm{ext}}^{U}  \) inject turbulent energy into the system and are described later in this section.
Because Eq.~\eqref{eqn:ohm} implies that the magnetic field
is frozen into the electron fluid, we include hyper-resistive dissipation,
\(-\eta_4 \nabla^4\bb{B}\), in Eq.~\eqref{eqn:faraday} to dissipate magnetic energy at scales far smaller than those at which ion heating dominantly occurs.
We therefore ascribe the energy that is dissipated by hyper-resistivity to be a proxy for electron heating (the feedback to the isothermal electron temperature is neglected).

We solve Eqs.~\eqref{eqn:hybrid-vlasov}--\eqref{eqn:ohm} using {\tt
  Pegasus++}
\citep{kunzPegasusNewHybridkinetic2014,arzamasskiyKineticTurbulenceCollisionless2023}, 
a massively
parallel, highly optimized,  particle-in-cell code that, following \citet{zhangExtremeHeatingMinor2025}, 
allows for multiple ion species.
The
plasma occupies a Cartesian, triply periodic, elongated box, with \((280)^2\times 1680\) cells. This box spans a domain of physical size
\(L_\perp^2\times L_\parallel = (28\pi\rho_{\rm p0})^2\times 168\pi\rho_{\rm p0}\), so that \(L_{\parallel}  = 6 L_{\perp}\), where \(L_{\perp }\) and \( L_{\parallel }\) are the box sizes perpendicular and parallel to \(\bb{B}_{0}\), respectively. These parameters imply a maximum resolved perpendicular wavenumber
of \(k_{\perp } \rho _{\mathrm{p}0}  = 10 \), minimum parallel and perpendicular
wavenumbers of \(k_{\parallel ,\min } \rho _{\mathrm{p}0} \approx  0.01 \) and
\(k_{\perp , \min } \rho _{\mathrm{p}0} \approx  0.07  \), and an Alfv\'en crossing
time
\(\tau _{\mathrm{A}} = L_{\parallel} / v_{\rm A} \approx 528\beta_{\rm p0}^{1/2}\,\Omega^{-1}_{\rm p}\).

The terms \(\bb{E}^{U} _{\rm ext} \) and
\(\grad \btimes \bb{E}^{B} _{\rm ext} \) in
Eqs.~\eqref{eqn:hybrid-vlasov} and \eqref{eqn:faraday} drive Alfv\'enically polarized fluctuations in
\(\bb{u}_{\perp } \) and \(\bb{B}_{\perp } \) at the outer scale of the box, as a proxy for the effect of larger, unresolved scales in the true solar wind. It is useful to express the fluctuations in terms of the Elsasser fields,
\begin{equation}\label{eq:elsasser}
\bb{z}^\pm \doteq \bb{u}_\perp \pm \frac{\bb{B}_\perp}{(4\pi m_{\rm p} n_{\rm p0})^{1/2}} \doteq \bb{u}_\perp\pm\bb{b}_\perp,
\end{equation}
which describe Alfv\'enic perturbations perpendicular (``\(\perp\)'') to the guide field that propagate in the \(\mp\hat{\bb{b}}\) direction, where \(\hat{\bb{b}}=\bb{B}/|\bb{B}|\).
We quantify the energy imbalance between the Elsasser fields using the normalized cross helicity,
\begin{equation}
\label{eq:sigmac}
    \sigma_{\rm c} \doteq \frac{\langle|\bb{z}^+|^2\rangle-\langle|\bb{z}^-|^2\rangle}{\langle|\bb{z}^+|^2\rangle+\langle|\bb{z}^-|^2\rangle} = \frac{2\langle\bb{u}_\perp\bcdot\bb{b}_\perp\rangle}{\langle|\bb{u}_\perp|^2\rangle+\langle|\bb{b}_\perp|^2\rangle},
\end{equation}
where $\langle\,\cdot\,\rangle$ denotes a box average.

The magnitudes of \(\bb{E}^{U} _{\rm ext} \) and
\(\grad \btimes \bb{E}^{B} _{\rm ext} \) are determined by specifying the rates at which energy \(\varepsilon  \) and cross helicity \(\varepsilon^{H} \) are injected into the Elsasser
fields \(\bb{z}^{\pm } \).
At each timestep, the spatial profiles of \(\bb{E}^{U} _{\rm ext}(t,\bb{r}) \) and
\(\grad \btimes \bb{E}^{B} _{\rm ext} (t,\bb{r})\) are sums of
Fourier modes with wavenumbers \(k_{j} \) satisfying
\({2\pi / L_{j} \leq k_{j}  \leq  4\pi  / L_{j} }  \), where \(L_{j} \) is the box length in direction \(j \in \{x,y,z\}\). Each mode is divergence-free and oriented perpendicular
to \(\bb{B}_{0} \). The Fourier
coefficients of these modes evolve in time via an
Ornstein--Uhlenbeck process with correlation time \(\tau _{\mathrm{corr}} \), with mode energy
normalized by \(k^{-2}\). They are normalized such that
\(  \langle m_{\rm p} n_{{\rm p}0}  \Omega_{\mathrm{p}}^{-1} |\bb{F}^{\pm }|^{2} \rangle = \varepsilon^{\pm }   \), where
\(\varepsilon^{\pm } = (\varepsilon \pm \varepsilon^{H} ) /2    \) are the desired energy injection rates into
each of the Elsasser fields, and \(\bb{F}^{\pm} \doteq q_{\rm p} \bb{E}^{U} _{\rm ext}/m_{\rm p} \pm c \grad \btimes \bb{E}^{B} _{\rm ext} / (4\pi m_{\rm p} n_{{\rm p}0})^{1/2} \). Because the forcing is time correlated, the
actual injection of energy (\(\varepsilon_{\mathrm{inj}}= \sum_{\pm} m_{\rm p} n_{{\rm p}0} \langle \bb{z}^{\pm }\bcdot\bb{F}^{\pm }\rangle /2 \)) and imbalance (\(\varepsilon^{H}  _{\mathrm{inj}} / \varepsilon_{\mathrm{inj}}   \)) can vary in time, and are generally not equal to the specified values of \(\varepsilon \) and \(\varepsilon ^{H}  \).\footnote{Given the
qualitative agreement in the development of the helicity barrier and subsequent
proton cyclotron heating between our \(\beta _{\mathrm{p} 0} = 0.3 \) imbalanced
simulation \citep{zhangExtremeHeatingMinor2025} and a previous simulation using
a different forcing method that held 
\(\varepsilon _{\rm inj} \) and \(\varepsilon^{H}  _{\rm inj} \) fixed
\citep{squireHighfrequencyHeatingSolar2022}, we expect our conclusions
to be robust to the details of the forcing.}

We perform six simulations that together form a parameter study of minor-ion heating under conditions relevant to different regions of the solar corona and solar wind.
Each simulation is characterized by a value of 
\(\beta _{\mathrm{p}0} \in \{1,0.3,1 /16\} \) and drives turbulence with either \(\varepsilon _{H} / \varepsilon  \in \{ 0,0.9\} \). We choose \(\varepsilon \approx (2\times 10^{-5}) m_{\rm p} v_{{\rm A}0}^{2} \Omega_{\rm p} / 2\mathcal{V}\) so that, if outer-scale fluctuations are critically balanced with correlation scales comparable to the elongated box, \(\delta B_{\perp}/B_{0} \approx L_{\perp } / L_{\parallel }\), then fluctuations at \(k_{\perp}\rho_{\mathrm{p}0}\approx1\) have spectral anisotropies comparable to turbulent eddies measured at similar scales in the solar wind \citep{chenRecentProgressAstrophysical2016a}.
Driving turbulence at low plasma beta is more difficult because of the reduced scale separation between the forcing scales and the ion-inertial length. To compensate, we increase $ \varepsilon$ to \({\approx}(6.3\times10^{-5})m_{\rm p}v_{{\rm A}0}^{2}\Omega_{\rm p}/2\mathcal{V}\) for the runs at \(\beta_{\mathrm{p}0}=1/16\).
In general, \(\varepsilon_{\mathrm{inj}}\) varies on the order of the forcing correlation timescale, \(\tau_{\mathrm{corr}}\).
Decreasing \(\tau_{\mathrm{corr}}\) reduces this temporal variation, but \(\tau_{\mathrm{corr}}\) cannot be made too small without directly driving fluctuations near ion gyrofrequencies.
To balance these requirements, we choose \(\tau_{\mathrm{corr}}=\tau_{\mathrm{A}}/2\) for the two simulations at \(\beta_{\mathrm{p}0}=1/16\).
Because the higher-beta simulations already have greater separation between forcing and ion-gyroscale frequencies, corresponding to larger \(\tau_{\rm A}\Omega_{\rm p}\), we instead choose \(\tau_{\mathrm{corr}}=\tau_{\mathrm{A}}/4\) for those runs to further limit temporal variations in \(\varepsilon_{\mathrm{inj}}\).
In all simulations, we use \(N_{\mathrm{ppc}} =1000\)
proton macroparticles per cell to alleviate concerns of electric-field noise and
numerical cooling
\citep{squireHighfrequencyHeatingSolar2022}.
To isolate the effects of imbalance
and \(\beta _{\mathrm{p} 0} \) on ion heating from our choices of minor-ion
species and abundances, we treat all minor-ion species
passively, \(\Lambda_i = 0\) for \(i \ne {\rm p}\).
Because these passive minor ions do not feed back and thereby do not contribute electromagnetic noise, we are able to use far fewer particles per
cell for each minor-ion species. We use \(N_{\mathrm{ppc}}=64\) for minor ions in the \(\beta_{\mathrm{p}0}=0.3\) imbalanced run and \(N_{\mathrm{ppc}}=27\) per minor-ion species in the other simulations. Peculiar velocity \(\bb{w}\) is defined with respect to the bulk proton flow, \(\bb{u}_{\rm p} = n_{{\rm p}}^{-1} \int \rmd \bb{r} \,\bb{v} f_{{\rm p}} \), such that \(\bb{w}=\bb{v}-\bb{u}_{\rm p}\).

For most simulations, in addition to bulk protons (\({\mathrm p}\)), we include alpha particles (\(\alpha \), \(\mathrm{He}^{2+}\)); quintuply and sextuply ionized oxygen, \({\mathrm{O}}^{5+} \) and \({\mathrm{O}}^{6+} \); quintuply and sextuply
ionized carbon, \({\mathrm{C}}^{5+} \) and \({\mathrm{C}}^{6+} \); and
nonuply ionized magnesium, \({\mathrm{Mg}}^{9+} \).
The exceptions are the \(\beta _{\mathrm{p}0} =0.3 \) imbalanced run, which includes only \(\alpha \) and \({\mathrm{O}}^{5+} \), and the \(\beta _{\mathrm{p}0} =0.3 \) balanced run, which includes \({\mathrm{Fe}}^{9+} \) instead of \({\mathrm{Mg}}^{9+} \) 
(because of its relatively large Larmor scale, \({\mathrm{Fe}}^{9+} \) suffers from direct heating by the forcing and so should be treated cautiously).
These minor-ion species are chosen for their presence in the
solar wind
\citep{bochslerAbundancesChargeStates2000,tracyConstrainingSolarWind2016}, and
to allow comparisons between ions that share the same mass, charge, or
mass-to-charge ratio.
The species \({\mathrm{O}}^{6+} \) is the third-most abundant ion and the
focus of recent minor-ion temperature measurements from Solar Orbiter
\citep{liviFirstResultsSolar2023a,riveraObservationalConstraintsRadial2025}.
Although less abundant, \({\mathrm{O}}^{5+} \) is observable in
remote-sensing \emph{UVCS} observations of the extended solar corona, where it exhibits extreme heating \citep{kohlUltravioletSpectroscopyExtended2006}.
Alphas, the second-most abundant ion species in the solar wind with fractional abundances of \({\sim}1\%{-}10\%\), are well diagnosed \citep{kasperSolarWindHelium2007a,marschHeliosEvolutionDistribution2012,verscharenMultiscaleNatureSolar2019}.
Ongoing PSP measurements probe their temperature and temperature
anisotropy in near-Sun regions where minor-ion heating is expected to be strongest
\citep{mostafaviPreferentialEnergizationSolar2025,mostafaviParkerSolarProbe2024a}.
Their abundance and high kinetic-energy content
\citep{wangAlphaProtonRelative2025} make the passive treatment of alphas less justified. However, a linear analysis of the \(\beta _{\mathrm{p} 0} =0.3 \)
imbalanced run suggests that their active feedback on heating would be minimal \citep{zhangExtremeHeatingMinor2025}.

\begin{deluxetable*}{ccccccc}
\tablecaption{Properties of the six simulations time averaged from when \(z_{\rm rms}^{+}\) saturates. Columns report the initial proton beta, input values of \(\varepsilon_{\mathrm{H}}/\varepsilon\), cross helicity, magnetic-fluctuation amplitude, measured energy injection rate $\varepsilon_{\mathrm{inj,rms}}$ in units of $m_{\rm p} v_{{\rm A}0}^{2} \Omega_{\rm p}/2\mathcal{V}$, $1-\varepsilon_{\mathrm{inj,H,rms}}/\varepsilon_{\mathrm{inj,rms}}$, and the measured electron heating $ Q_{\mathrm{e}}/\mathcal{V}\varepsilon_{\mathrm{inj,rms}}$. The \(\beta_{{\rm p}0}=1/16\) balanced simulation has a lower late-time total dissipation,
\((Q_{\rm p}+Q_{\rm e})/\mathcal{V}\varepsilon_{\mathrm{inj,rms}}\approx0.47\),
than the other runs, for which
\((Q_{\rm p}+Q_{\rm e})/\mathcal{V}\varepsilon_{\mathrm{inj,rms}}\approx1\).
This reflects the short saturated interval available for time averaging, since this run was evolved only to \(t=3.8\tau_{\rm A}\).
\label{tab:results-summary}}
\tablehead{
\colhead{$\beta_{\mathrm{p}0}$} &
\colhead{$\varepsilon_{\mathrm{H}}/\varepsilon$} &
\colhead{$|\sigma_{\mathrm{c}}|$} &
\colhead{$\delta B_{\mathrm{rms}}/B_0$} &
\colhead{$\varepsilon_{\mathrm{inj,rms}}$} &
\colhead{$1-\varepsilon_{\mathrm{inj,H,rms}}/\varepsilon_{\mathrm{inj,rms}}$} &
\colhead{$ Q_{\mathrm{e}} /\mathcal{V}\varepsilon_{\mathrm{inj,rms}}$}
}
\startdata
1     & 0.9 & 0.97     & 0.52 & $9.3\times10^{-5}$ & 0.05 & 0.12 \\
1     & 0   & 0.23 & 0.33 & $2.0\times10^{-4}$ & N/A  & 0.60 \\
0.3   & 0.9 & 0.98     & 0.33 & $5.9\times10^{-5}$ & 0.08 & 0.07 \\
0.3   & 0   &  0.024 & 0.19 & $8.5\times10^{-5}$ & N/A  & 0.60 \\
1/16  & 0.9 & 0.93     & 0.15 & $3.2\times10^{-5}$ & 0.05 & 0.07 \\
1/16  & 0   & 0.21 & 0.15 & $1.7\times10^{-4}$ & N/A  & 0.20 \\
\enddata
\end{deluxetable*}

\section{Results}
\label{sec:results}

We now present our results.
Section~\ref{sec:time-evol-turb} focuses on the time evolution of turbulence and minor-ion heating in the simulations, while \S\ref{sec:mass-charge-scalings} compares the mass--charge dependence of measured minor-ion heating with the theoretical predictions of \S\ref{sec:theory}.
Throughout this section, we present results from the balanced and imbalanced simulations together, so that their qualitatively different evolution can be contrasted for each diagnostic.

\subsection{Time evolution of turbulence and minor-ion kinetics}
\label{sec:time-evol-turb}

\subsubsection{Reduced quantities}

\begin{figure*}
    \centering
    \includegraphics[width=0.95\textwidth]{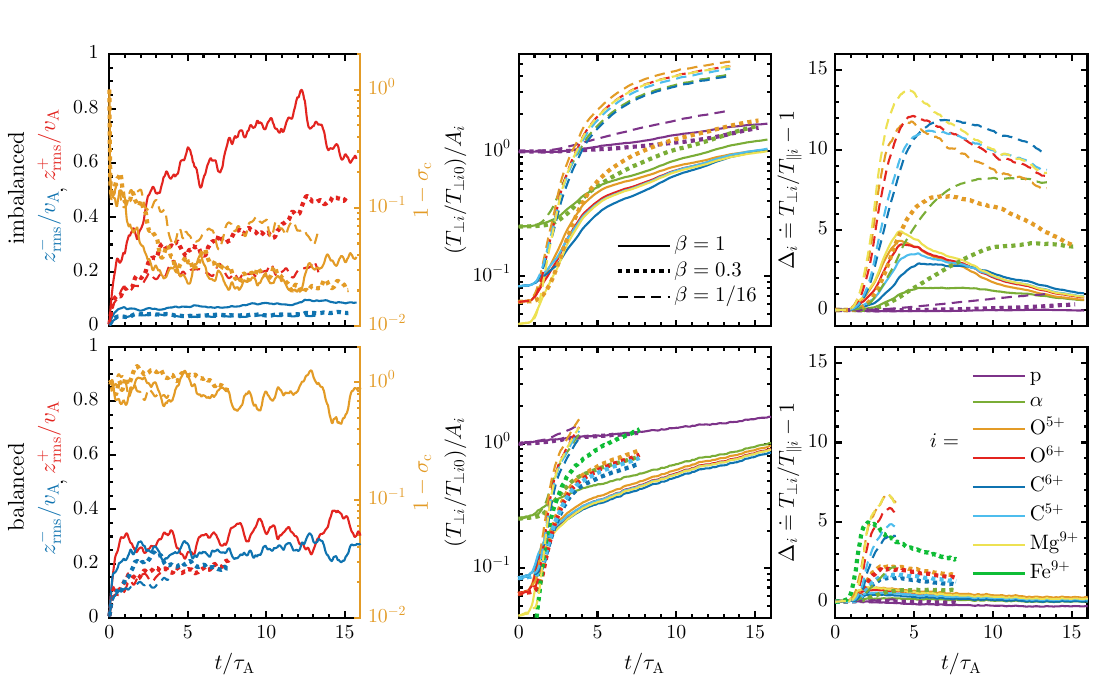}
    \caption{Left panel shows time evolution of $z^{\pm}_{\rm rms}$ in red/blue and $\sigma_{c}$
      in yellow (right axis). Middle and right panels show mass-normalized temperatures and temperature
      anisotropies, respectively, of protons and minor ions as indicated by colors on legend, throughout the imbalanced
      (top) and balanced (bottom) simulations at \(\beta = 1\) (solid), \(\beta = 0.3\)
      (dotted), and \(\beta = 1/ 16\) (dashed).}
    \label{fig:zandT}
\end{figure*}

The time evolution of the fluctuation amplitudes, \(z^{\pm}_{\mathrm{rms}}\doteq\langle|\bb{z}^{\pm}|^2\rangle^{1/2}\), and turbulence imbalance, \(\sigma_{\rm c}\), is shown in the left column of Fig.~\ref{fig:zandT}.
In the simulations with balanced driving, the fluctuation amplitudes saturate within a few turnover times, by \(2\tau_{\rm A}\).
This evolution differs starkly from the runs with imbalanced driving. 
In those runs, only the subdominant fluctuation amplitude, \(z_{\rm rms}^{-}\), saturates by \(2\tau_{\rm A}\).
The dominant fluctuation amplitude, \(z_{\rm rms}^{+}\), continues to grow and saturates later.
This evolution is consistent with the helicity barrier, which allows only the balanced portion of the cascade flux, \({\approx}2\varepsilon^{-}_{\rm inj}\), to reach small scales, \(k_{\perp}\rho_{\mathrm{p}}\gtrsim1\), and dissipate through hyper-resistivity.
Consequently, \(z_{\rm rms}^{-}\) saturates quickly, while \(z_{\rm rms}^{+}\), which carries the dominant imbalanced portion of the cascade, continues to grow because it cannot dissipate efficiently through small scales.
It saturates later only at larger amplitudes, when ion heating becomes sufficient to regulate the cascade.

The late-time values of \(\sigma_{\mathrm{c}}\), \(\delta B_{\mathrm{rms}}/B_0\), \(\varepsilon_{\mathrm{inj,rms}}\), \({1-\varepsilon_{\mathrm{inj,H,rms}}/\varepsilon_{\mathrm{inj,rms}}}\), and \( Q_{\mathrm{e}}/\mathcal{V}\varepsilon_{\mathrm{inj,rms}}\) for all six runs are summarized in Table~\ref{tab:results-summary}.
Despite the lower values of \(\varepsilon_{\mathrm{inj,rms}}\) in the runs with imbalanced driving, the magnetic fluctuation amplitudes \(\delta B_{\mathrm{rms}}/B_0\) are comparable to or larger than the amplitudes in the corresponding balanced runs.
The theory describing the helicity barrier predicts that only the balanced portion of the injected energy proceeds to scales satisfying \(k_{\perp}\rho_{\mathrm{p}}\gg1\), where it is ultimately dissipated as electron heating, so that \(Q_{\mathrm{e}}/\mathcal{V}\varepsilon_{\mathrm{inj}}\approx1-\varepsilon_{\mathrm{inj,H}}/\varepsilon_{\mathrm{inj}}\).
The measured values of \(1-\varepsilon_{\mathrm{inj,H}}/\varepsilon_{\mathrm{inj}}\) are consistent, particularly at lower beta, with the late-time-averaged electron heating rates \(Q_{\mathrm{e}}/\mathcal{V}\varepsilon_{\mathrm{inj}}\) in Table~\ref{tab:results-summary}.
These heating rates are inferred by subtracting the proton heating and the rate of change of bulk-field energy from \(\varepsilon_{\mathrm{inj}}\).
In the runs with balanced driving, \(\varepsilon_{\mathrm{H}}/\varepsilon=0\), for which we force only \(\bb{u}_{\perp}\), slight energy imbalances develop for \(\beta_{\mathrm{p}0}=1\) and \(\beta_{\mathrm{p}0}=1/16\).
Nevertheless, no signatures of a helicity barrier are seen in these balanced-driving simulations, leading to decreased proton-to-electron heating ratios relative to the imbalanced runs, as given in Table~\ref{tab:heating-ratios}.
In balanced turbulence, the enhanced proton heating at \(\beta_{\mathrm{p}0}=1/16\) compared to the higher-beta runs arises from stronger stochastic heating \citep{chandranPerpendicularIonHeating2010b,cerriStochasticHeatingIts2021a}, which is absent in gyrokinetic theories that predict the opposite trend with beta \citep{kawazuraThermalDisequilibrationIons2019,schekochihinConstraintsIonElectron2019}.

The evolution of minor-ion temperatures and temperature anisotropies is shown in the middle and right columns of Fig.~\ref{fig:zandT}.
Both quantities grow most strongly in imbalanced turbulence and at low \(\beta_{\mathrm{p}0}\).
The ions are initially isothermal with one another, so heavier ions begin
with lower 
mass-normalized temperatures.
In all simulations, minor-ion temperatures approach approximate mass proportionality with one another, with additional weaker dependence on both mass and charge.
In the \(\beta _{\mathrm{p}0}=1 /16 \) imbalanced
simulation, where heating is strongest, minor-ion temperatures surpass
mass-proportionality relative to the bulk protons, reaching
\(T_{\perp\mathrm{O}^{5+}}/T_{\perp\mathrm{p}}\approx39\) and \(T_{\perp\alpha}/T_{\perp\mathrm{p}}\approx7.8\) by the end of the simulation, compared with \(A_{\mathrm{O}^{5+}}=16\) and \(A_{\alpha}=4\). For heavy ions (those with \(A_i>4\)), temperature anisotropies become extreme, with \(T_{\perp i} / T_{\parallel i} \gtrsim 10  \), while \(T_{\perp \alpha } / T_{\parallel \alpha } \lesssim  8  \).
At \({\beta _{\mathrm{p}0}= 1 / 16 }\), the imbalanced simulation reproduces the extreme
\(\mathrm{O}^{5+} \) temperatures, \(T_{\perp  \mathrm{O}^{5+} } / T_{\perp  \mathrm{p}}  \gtrsim  40 \),
and anisotropies, \(T_{\perp {\mathrm{O}}^{5+} } / T_{\parallel {\mathrm{O}}^{5+} } \sim   10  \), inferred
from remote-sensing measurements of coronal holes
\citep{kohlUltravioletSpectroscopyExtended2006,cranmerImprovedConstraintsPreferential2008}. Likewise, the alpha-to-proton temperature ratio and perpendicular anisotropy are consistent with near-Sun solar wind  measurements \citep{mostafaviParkerSolarProbe2024a,mostafaviPreferentialEnergizationSolar2025}.

\subsubsection{Development of small parallel scales and PCWs}
\begin{figure*}
    \centering
    \includegraphics[width=0.95\textwidth]{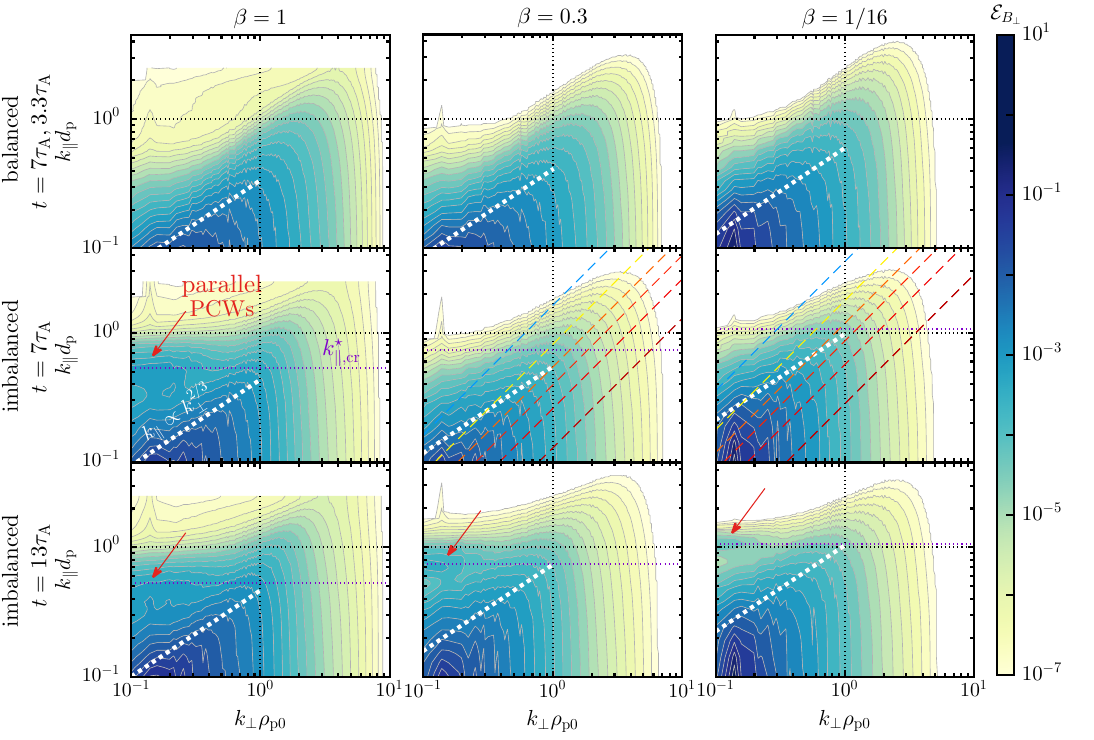}
    \caption{Spectra of perpendicular magnetic-field fluctuations,
      $\mathcal{E}_{B_{\perp}}= \mathcal{E}_{B_{y}} + \mathcal{E}_{B_{z}}$, in $(k_{\perp},k_{\parallel})$ space, with logarithmic
      contours and color bar in units of \(m_{{\mathrm p}0} n_{{\mathrm p}0} v_{{\mathrm A} 0}^2/2\), for the \(\beta = 1\) (left column), \(\beta =0.3\) (middle
      column), and \(\beta =1 /16\) (right column) imbalanced simulations at intermediate stages of the evolution
      (\(t=7\tau _{\mathrm{A}} \)) (middle row) and saturation \(t=13\tau _{\mathrm{A}} \) (bottom row).
      Equivalent spectra for the balanced simulations are shown at saturation in the top row (\(t=7\tau _{\mathrm{A}} \), for \(\beta =1\) and \(\beta =0.3\),
      and \(t =3.3\tau _{\mathrm{A}} \), for \(\beta =1 /16\)).
      White
      dotted lines, plotted visually, show an approximate flow of peak energy, i.e., the top edge of the critical-balance cone under which
      energy cascades to smaller scales; the black dotted lines mark
      $k_{\perp} \rho_{\rm p0}=1$ and $k_{\parallel} d_{\rm p0}=1$. The presence of parallel
      proton-cyclotron waves (PCWs) is visible from the enhanced power at larger
      \(k_{\parallel } \) and small \(k_{\perp }  \) at later times in the imbalanced
      simulations, and is also explicitly labeled at intermediate time for
      \(\beta =1\). The purple dotted lines indicate \(k_{\parallel,{\rm cr}}^{\star} d_{\rm p}\). The colored dashed lines show equal area angular bins in \(k_{\perp}\)-\(k_{\parallel}\) as described in the text.}
    \label{fig:2dspec}
\end{figure*}

A key element of our quasilinear theory of minor-ion heating is the magnetic power spectrum, shown in \(k_{\parallel }\)-\(k_{\perp }  \) space in
Fig.~\ref{fig:2dspec}. These spectra are computed using the field-line-following method described
by \citet{squireHighfrequencyHeatingSolar2022}.
As anticipated from the helicity barrier and reflected in Fig.~\ref{fig:zandT}, inertial-range power builds up with time because the dominant imbalanced portion of the cascade flux cannot proceed to \(k_{\perp}\rho_{\mathrm{p}0}\gtrsim1\).
This buildup is visible in Fig.~\ref{fig:2dspec} as enhanced power at higher \(k_{\parallel}\) in the imbalanced simulations (middle and bottom rows) relative to the saturated balanced simulations (top row), even at intermediate times, \(t\approx7\tau_{\mathrm{A}}\).
Across the columns of Fig.~\ref{fig:2dspec}, decreasing \(\beta_{\mathrm{p}0}\) shifts the power at fixed \(k_{\perp}\rho_{\mathrm{p}0}\) to higher \(k_{\parallel}d_{\mathrm{p}0}\).
This trend is reflected by the upward shift of the white dotted lines,  which provide guides to the flow of peak fluctuation power below the \(k_{\parallel}\propto k_{\perp}^{2/3}\) critical-balance curve.
In the imbalanced simulations, the enhanced fluctuation power at high \(k_{\parallel}\) promotes cyclotron-resonant ion heating and produces the strongest temperature growth seen in Fig.~\ref{fig:zandT}.

Another feature of the imbalanced spectra in Fig.~\ref{fig:2dspec} is the appearance of parallel PCWs, indicated
 by red arrows, which arise from quasilinear focusing
(\S\ref{sec:ql-focusing}).
These parallel PCWs grow earlier and at lower \(k_{\parallel }d_{\mathrm{p}0}  \) for 
higher \(\beta _{\mathrm{p}0} \). 
Although quasilinear theory predicts that the parallel scales of the parallel PCWs and the oblique PCWs at \(k_{\perp}^{\star}\) differ (\(k_{\parallel,\mathrm{P}}^{\star}\) and \(k_{\parallel,\mathrm{cr}}^{\star}\), respectively) \citep{yergerCyclotronBreakingMechanism2026}, the difference is small enough in our simulations to be unmeasurable at late times.
At higher \(\beta_{\mathrm{p}0}\), the resonance condition for protons with \(w_{\parallel}\sim v_{{\rm th,p}}\) is satisfied at lower \(k_{\parallel}\). As a result, a larger fraction of the turbulent cascade flux can be quasilinearly focused into parallel PCWs earlier in the evolution.
If these PCWs influence minor-ion heating through the mechanism described in \S\ref{sec:ql-focusing}, their effect should be stronger at higher \(\beta_{\mathrm{p}0}\), where they reach larger amplitudes and their lower \(k_{\parallel,\mathrm{P}}^{\star}d_{\mathrm{p}0}\) resonates with a larger fraction of the minor-ion population.

\subsubsection{Anisotropy of non-Maxwellian VDFs}
\begin{figure*}
    \centering
    \includegraphics[width=0.95\textwidth]{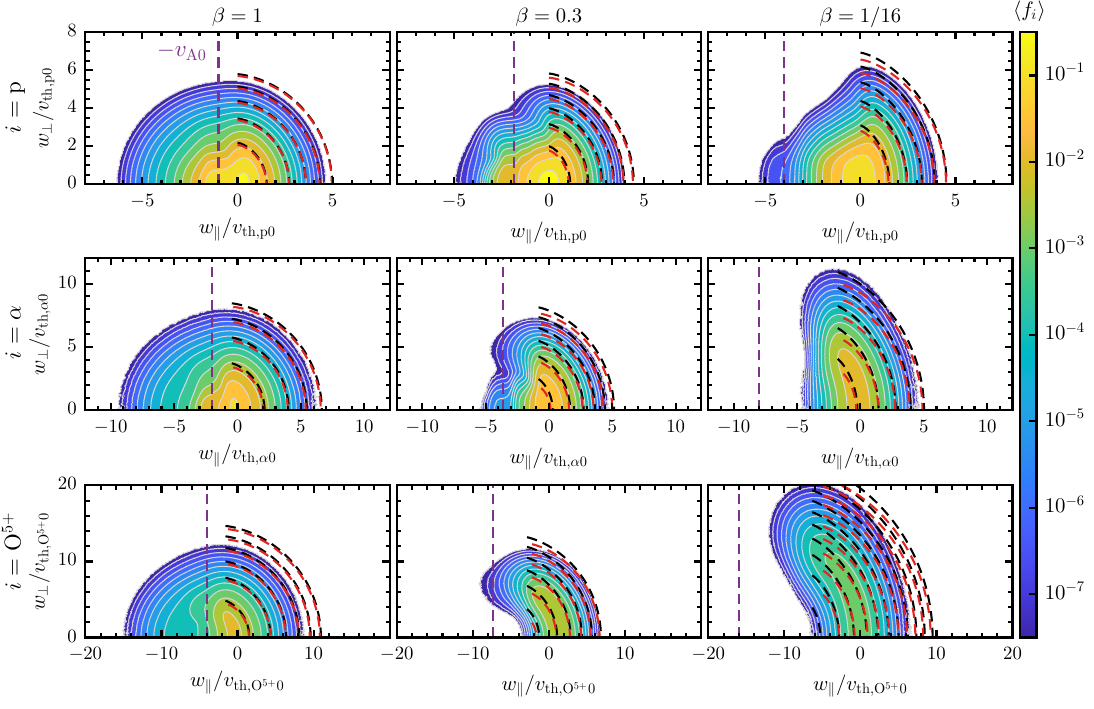}
    \caption{Imbalanced simulation VDFs, \(f_i(w_\parallel,w_\perp)\), in peculiar velocity
    \((w_{\parallel}, w_{\perp})\) of protons (top row), alphas (middle row),
    and O\(^{5+}\) (bottom row) at \(t=13\tau _{\mathrm{A}} \) for  \(\beta_{\mathrm{p}0}  =1\) (left column),
    \(\beta _{\mathrm{p}0}= 0.3 \) (middle column), and
    \(\beta _{\mathrm{p}0} = 1 / 16 \) (right column), with logarithmic color bar
    and contours. The axes are scaled to the initial thermal speeds of each
    respective species, \(\vth{,i0}\), and \(w_{\parallel } \) is taken with respect to
    the bulk plasma (proton) flow. The kinetic-Alfv\'en-wave Landau resonance is indicated by the vertical purple dashed line.
For each species, the black and red dashed lines trace the \(n=1\) cyclotron-resonant scattering contours for oblique and parallel PCWs, respectively, propagating in the \(-\hat{\bb{b}}\) direction.}
    \label{fig:vdfs}
\end{figure*}

Cyclotron-resonant heating by oblique PCWs is the dominant ion-heating mechanism in the imbalanced simulations.
Quasilinear theory accurately predicts the resulting evolution towards highly asymmetric, non-bi-Maxwellian ion VDFs. The ion VDFs are plotted in gyrotropic peculiar-velocity space \((w_{\parallel},w_{\perp}) \) in
Fig.~\ref{fig:vdfs}, with the axes normalized to the initial thermal speed of each ion species, \((w_{\parallel}/v_{{\rm th},i0},w_{\perp}/v_{{\rm th},i0})\); different axis ranges are chosen to show the final VDF structure more clearly. All VDFs flatten along oblique-PCW scattering contours (black dashed lines) 
and across Landau resonances (vertical purple dashed lines), with the minor-ion VDFs flattening along oblique-PCW contours that extend into \(w_{\parallel}<0\) up to \(w_{\parallel,\min} \). The
mechanism of quasilinear focusing, described in \S\ref{sec:ql-focusing},
is illustrated by the red, dashed curves that trace the parallel-PCW scattering contours; these contours are shallower than both the oblique-PCW contours and the isocontours of the quasilinearly saturated ion VDFs.
Throughout most of velocity space, the VDF gradients are oriented such that diffusion along the parallel-PCW contours cools the ions; for the active bulk protons, \(\Lambda_{\rm p}=1\), this diffusion also emits parallel PCWs.
This diffusion facilitates further heating by moving particles across saturated oblique-PCW contours onto contours residing at higher kinetic energies.

Many features of the time evolution of \(T_{\perp i}\) and \(\Delta_i=T_{\perp i} / T_{\parallel i} -1\) in Fig.~\ref{fig:zandT}, including their dependence on \(Z_i\) and \(A_i\), can be understood from the contour shapes in Figs.~\ref{fig:res-cont} and~\ref{fig:vdfs}.
During the early stages of heavy-ion heating, the scattering contours are nearly vertical in \(w_{\perp}\) relative to the initial VDFs, before curving toward the \(-w_{\parallel } \) direction.
The initial thermal speeds determine the size of the initially isothermal Maxwellian VDFs relative to the velocity-space scale over which the contours begin to curve.
This is apparent in Fig.~\ref{fig:res-cont}, where the axes are normalized to \(v_{\rm A}\) and the shaded regions, of approximate size \(|w_{\parallel}|<v_{{\rm th},i}\), indicate the dense initial VDF core of each species.
Although the \({\rm O}^{5+}\) scattering contours are only slightly steeper than the alpha contours, they extend roughly twice as far in \(w_{\perp}\) before curving relative to the slower \({\rm O}^{5+}\) thermal core.
Because heavier ions are initially slower, \(v_{{\rm th},i0}\propto A_i^{-1/2}\), their smaller VDFs are stretched farther in \(w_{\perp}\) before flattening in \(w_{\parallel}\).
The early-time peak temperature anisotropies in Fig.~\ref{fig:zandT} are therefore organized primarily by ion mass.

At late times, quasilinearly saturated ion VDFs  adopt the shapes of their resonant scattering
contours, which then determine the temperature anisotropy of each species. 
As
discussed in \S\ref{sec:theory} and shown in Fig.~\ref{fig:res-cont}, minor ions with larger
mass-to-charge ratios have scattering contours that are steeper in \(w_{\perp } \),
but extend significantly further into \(w_{\parallel } < 0\).
Because the interspecies differences in the extent of the contours, set by \(w_{\parallel,\min}\), dominate over differences in contour steepness, the late-time temperature anisotropies in Fig.~\ref{fig:zandT} correlate with decreasing ion mass-to-charge ratio. This contrasts with the early-time anisotropies, which depend primarily on ion mass.

The temperature-anisotropy trends predicted by this discussion are clearest for heavy ions in the imbalanced \(\beta_{\mathrm{p}0}=1/16\) simulation, whose initial VDFs occupy smaller regions of
velocity space and are fully covered by the scattering contours.
For example,
the early-time peaks favor larger ion mass, with \({\rm max}(\Delta_{\mathrm{Mg}^{9+}}) > {\rm max}(\Delta_{\mathrm{O}^{6+}}) \),
whereas the late-time anisotropies depend mainly on mass-to-charge ratio  and increase as \(A_i/Z_i\) decreases, as seen from  \(\Delta _{{\mathrm{C}^{6+} }} > \Delta _{{\mathrm{C}^{5+} }} \) and
\(\Delta _{{\mathrm{O}^{6+} }} > \Delta _{{\mathrm{O}^{5+} }} \).
Alphas are
lighter and have larger initial thermal speeds, so their initial VDFs are larger
relative to the scattering contours, leaving a substantial portion at
\(w_{\parallel } < w_{\parallel ,\min }  \) out of resonance.
Consequently, alphas do not
develop an early peak in temperature anisotropy, and, at late times, do not reach anisotropies as large as those of \(\mathrm{C}^{6+} \), despite having the same gyrofrequency.
At higher
\(\beta _{\mathrm{p}0} \), faster ion thermal speeds make the initial ion VDFs
broader relative to the scattering contours and place more particles with \(w_{\parallel}<0\) out of resonance. Together with increased Landau damping, which
exacerbates the non-bi-Maxwellian curvature into \(w_{\parallel }<0 \), this reduces ion temperature anisotropies at higher plasma beta. Overall, the evolution of ion temperature anisotropy can be understood from the shapes of the scattering contours relative to the initial ion VDFs: peak anisotropies are organized primarily by ion mass, saturated anisotropies are organized primarily by mass-to-charge ratio, and anisotropies are generally larger at lower \(\beta_{\mathrm{p}0}\).

The VDFs in simulations of balanced turbulence (not shown) do not indicate
quasilinear flattening along resonant contours and exhibit little asymmetry in
\(w_{\parallel } \).
This lack of quasilinear flattening is expected because nonlinear frequency broadening is stronger in balanced turbulence, making resonant-heating phenomenologies less applicable \citep{johnstonQuasilinearTheoryPerpendicular2025}.
Instead, the reduced distributions, \(f_i(w_{\perp } )\), are flat-topped for all ion species in balanced turbulence \citep[e.g., the \(\beta_{\mathrm{p}0}  =0.3\) balanced simulation in
the bottom panel of figure 7 of][]{zhangExtremeHeatingMinor2025}, consistent with predictions for stochastic heating \citep{kleinEvolutionProtonVelocity2016}.
Regardless, wave--particle interactions, including Landau damping, which in imbalanced turbulence produce \(w_{\parallel}\) asymmetries and drifts, instead contribute primarily to increasing the parallel temperature when co- and counter-propagating fluctuations are balanced. 
The weaker temperature anisotropies seen in Fig.~\ref{fig:zandT} for the balanced-driving simulations are therefore expected.

\subsection{Mass-charge dependence of minor-ion heating}
\label{sec:mass-charge-scalings}

The broad set of minor-ion species in our simulations provides detailed
empirical constraints on the mass--charge dependence of ion heating.
Although the minor-ion temperatures in Fig.~\ref{fig:zandT} are approximately mass proportional, additional dependencies on mass and charge accumulate over the heating history.
For direct comparison with the quasilinear theory in \S\ref{sec:ql-minor-heat}, we use the non-dispersive limit of Eq.~\eqref{eq:ql-heat-rate-disp}, which predicts a simple power-law dependence of minor-ion heating rates on mass-to-charge ratio,
\(Q_i\propto A_i(A_i/Z_i)^{a_{\mathrm{pred}}}\).
The prediction is \(a_{\mathrm{pred}}=\eta-2\), where \(\mathcal{E}_{E_{\perp}}\sim k_{\parallel}^{-\eta}\) is the perpendicular electric-field energy spectrum in \(k_{\parallel}\) of the resonant wave mode.
We calculate heating rates in the simulations from the rate of change of the box-integrated temperature
of each
species,
\begin{gather}
Q_{\perp i}\doteq \dv{}{t} \int \rmd{\bb{r}}\, \int \rmd{\bb{w}}\,   \frac{1}{2} m_i w^2_\perp f_i , \label{eqn:Qprp}\\*
Q_{\parallel i}\doteq \dv{}{t} \int \rmd{\bb{r}}\, \int \rmd{\bb{w}}\, \frac{1}{2} m_i w^2_\parallel f_i, \label{eqn:Qprl}
\end{gather}
with \(Q_i = Q_{\perp  i} + Q_{\parallel  i}  \).
We infer the time-dependent power-law index \(a(t)\) describing the dependence of the heating rate on mass-to-charge ratio by least-squares regression of the linear relationship,
\(\log (Q_i (t)/ A_i) = a(t) \log (A_i / Z_i) + \log  K(t) \), with \(K(t)\) the constant of
proportionality of the power law.
At each time, the fit uses only the instantaneous minor-ion heating rates.

\begin{figure}[htbp]
    \centering
    \includegraphics[width=0.95\columnwidth]{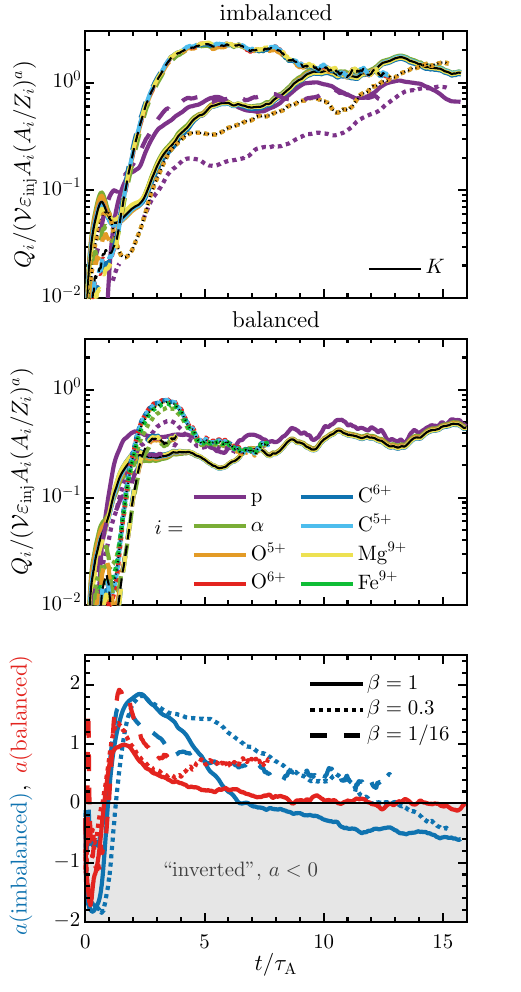}
    \caption{Top and middle:
    Total heating rates for each ion species (different colors; see legend), normalized by the box volume, time-averaged energy injection, and the time-dependent fits from the bottom panel, in the imbalanced (top) and balanced (middle) simulations. The thinner black lines show the fitted values of the constant of proportionality, \(K\). Because the fitted power laws collapse the minor-ion heating rates, many of the colored curves lie nearly on top of one another and are not individually visible.
      Bottom: Time evolution of the power-law index, \(a\), obtained by fitting  quantities proportional to (\(A_i / Z_i) ^{a} \) for the imbalanced (blue) and
      balanced (red) simulations at \(\beta _{\mathrm{p}0} = 1 \) (solid),
      \(\beta _{\mathrm{p}0} = 0.3 \) (dotted), and \(\beta _{\mathrm{p}0} = 1 /16 \)
      (dashed).  The
      fitted quantities are the heating rates normalized by averaged energy
      injection, volume, and ion mass. The shaded region marks \(a<0\), corresponding to an inverted dependence of heating rate on mass-to-charge ratio.
      }
    \label{fig:mqscaling}
\end{figure}

Fig.~\ref{fig:mqscaling} demonstrates that the minor-ion heating rates in our simulations are fit extremely well by empirical 
power-laws in ion mass-to-charge ratio. The first two rows of the figure show the heating rate of each species, normalized by the box volume, the late-time-averaged energy injection rate, and the fitted mass--charge dependence, \(A_i(A_i/Z_i)^{a(t)}\), with the color of the lines indicating the species type. With this fit and normalization, the minor-ion heating rates collapse onto one another.
The black lines show the fitted proportionality constant, \(K(t)\), which measures the efficiency of
minor-ion heating relative to the energy injection rate after removing the mass--charge
dependence.

For passive minor ions,
\(Q_i (t) / ( \mathcal{V}\varepsilon _{\mathrm{inj}}) \)
and  \(K(t)\) can exceed unity.
This occurs, for example, in the imbalanced simulations shown in the top panel of Fig.~\ref{fig:mqscaling}, where
\( \langle K \rangle_{t}   \gtrsim  1{-}1.5\) by saturation.
Because our minor ions are passive and do not remove energy from the electromagnetic fields, these heating rates measure how rapidly the temperature of each species changes, rather than the fraction of cascade energy absorbed by that species.
The large values of \(K(t)\) reflect two effects in imbalanced turbulence: (i) enhanced ion heating relative to electron heating, corresponding to larger \(Q_{\mathrm{p}}/Q_{\mathrm{e}}\), and (ii) enhanced minor-ion heating relative to proton heating, corresponding to larger \(Q_i/Q_{\mathrm{p}}\).
Both effects arise naturally from the helicity barrier, as follows.
\begin{enumerate}
\renewcommand{\labelenumi}{(\roman{enumi})}
\item
\emph{Enhancement of \(Q_{\mathrm{p}}/Q_{\rm e}\).}
Because of the helicity barrier, only the subdominant balanced portion of the cascade flux reaches small scales and heats electrons (\S\ref{sec:time-evol-turb} and Table~\ref{tab:results-summary}).
In saturation, the dominant imbalanced portion must instead dissipate through ion heating.
This is apparent in Fig.~\ref{fig:mqscaling}: the purple lines, corresponding to proton heating, compose a larger fraction of \(\varepsilon_{\mathrm{inj}}\) in the imbalanced runs in the top row than in the balanced runs in the middle row.
The saturated proton-to-electron heating ratios, given for each simulation in Table~\ref{tab:heating-ratios}, are larger by factors of \({\approx}6{-}30\) when the driving is imbalanced.
\item
\emph{Enhancement of \(Q_i/Q_{\mathrm{p}}\).}
This enhancement arises from both wave dispersion and the helicity barrier, because protons resonate with fluctuations at particularly high \(k_{\parallel}\), where the power falls off steeply.
The fluctuation power that heats the core of the proton distribution, \(w_{\parallel}\lesssim v_{{\rm th,p}}\), lies at \(k_{\parallel}>k_{\parallel,\mathrm{cr}}^{\star}\), with its local maximum near the transition-range cutoff at  \(k_{\perp}^{\star}\rho_{\mathrm{p}}\sim1\).
At \(k_{\perp}>k_{\perp}^{\star}\), the fluctuation power rapidly decreases through the transition range as a consequence of the helicity barrier.
By contrast, the \(k_{\parallel}<k_{\parallel,\mathrm{cr}}^{\star}\) fluctuations that resonate strongly with minor ions have maximum power at \(k_{\perp}\) well within the inertial range.
After integrating over \(k_{\perp}\), the \(k_{\parallel}\) spectrum steepens as \(k_{\parallel}\) approaches and exceeds \(k_{\parallel,\mathrm{cr}}^{\star}\), because an increasing fraction of the contributing power comes from fluctuations near or beyond the transition-range cutoff.
Thus, protons sample a steeper, lower-power portion of \(\mathcal{E}_{E_{\perp}}(k_{\parallel})\) than the minor ions, whose heating is controlled by the shallower \(\mathcal{E}_{E_{\perp}}\sim k_{\parallel}^{-\eta}\) scaling over their resonant range.
Dispersive effects, described by Eq.~\eqref{eq:ql-heat-rate-fullO}, further strengthen this distinction for protons: as \(w_{\parallel}\to0\), protons resonate with fluctuations at \(k_{\parallel,\mathrm{res}}(\Omega_{\rm p},w_{\parallel}=0)\to\infty\), where the power vanishes.
Thus, beyond the expected mass--charge dependence, minor ions are heated more strongly than protons, with \(K/Q_{\rm p}\approx1.7\) in imbalanced turbulence compared to \(K/Q_{\rm p}\sim1\) in balanced turbulence, as summarized in Table~\ref{tab:heating-ratios}.
\end{enumerate}

\begin{deluxetable}{ccccccc}
\tablecaption{Late-time heating ratios, \(Q_{\rm p}/Q_{\rm e}\) and \(K/Q_{\rm p}\), for the six simulations, averaged over the saturated state. Empirical power-law indices are given for the fits at \(t=7 \tau_{\rm A}\) (\(a_{7\tau_{\rm A}}\)), the end of each simulation (\(a_{\rm end}\)), and at times for each simulation where \(a\) is maximized (\(a_{\max}\)). \label{tab:heating-ratios}}
\tablehead{
\colhead{$\beta_{\mathrm{p}0}$} &
\colhead{$\varepsilon_{\mathrm{H}}/\varepsilon$} &
\colhead{$Q_{\mathrm{p}}/Q_{\mathrm{e}}$} &
\colhead{$K/Q_{\mathrm{p}}$} &
\colhead{$a_{\rm max}$} &
\colhead{$a_{7\tau_{\rm A}}$} &
\colhead{$a_{\rm end}$}
}
\startdata
1    & 0.9 & 4.81 & 1.80 & 1.8 & 0 & -0.6 \\
1    & 0   & 0.71 & 0.90 & 1 & 0.1 & 0 \\
0.3  & 0.9 & 12.9 & 1.70 & 1.8 & 1 & -0.4\\
0.3  & 0   & 0.47 & 1.00 & 1.3 & 0.75 & 0.7\\
1/16 & 0.9 & 10.2 & 1.66 & 1.6 & 0.6 & 0.5\\
1/16 & 0   & 1.37 & 1.27 & 1.9 & N/A & 0.75\\
\enddata
\end{deluxetable}

We now focus on the empirical mass--charge scalings described by the
fitted power-law indices
\(a(t,\sigma_{\mathrm{c}},\beta _{\mathrm{p} 0}  )\) in the bottom panel of Fig.~\ref{fig:mqscaling}, and assess whether the prediction of quasilinear theory for
\(a\) based on the slope of the electric-field spectrum is accurate.
In all simulations, a strong dependence of the form \(Q_i\sim A_i(A_i/Z_i)^a\), with \(1<a<2\), emerges at early times, \({\sim}1{-}2\tau_{\mathrm{A}}\), after the turbulent cascade develops.
Thereafter, the power law weakens and \(a\)
decreases in time.
For balanced turbulence, the power-law index saturates at \(a\approx0.75\) in the lower-beta simulations, \(\beta_{\mathrm{p}0}=0.3\) and \(\beta_{\mathrm{p}0}=1/16\), and reaches \(a\approx0\), corresponding to mass-proportional heating, at \(\beta_{\mathrm{p}0}=1\).
For imbalanced turbulence, the power-law index saturates only for \(\beta_{\mathrm{p}0}=1/16\), reaching \(a\approx0.5\), but continues to decrease into negative values for \(\beta_{\mathrm{p}0}=0.3\) and \(\beta_{\mathrm{p}0}=1\).
The gray shading in Fig.~\ref{fig:mqscaling} marks the corresponding ``inverted'' regime, \(a<0\), in which the heating-rate scaling decreases with \(A_i/Z_i\).
This inversion occurs earlier for \(\beta_{\mathrm{p}0}=1\), after \(t\approx7\tau_{\mathrm{A}}\), than for \(\beta_{\mathrm{p}0}=0.3\), where \(a<0\) after \(t\approx12\tau_{\mathrm{A}}\).
The early-time peak values, \(a_{\rm max}\), and final values, \(a_{\rm end}\), are summarized in Table~\ref{tab:heating-ratios}.

\begin{figure*}
    \centering
    \includegraphics[width=0.95\textwidth]{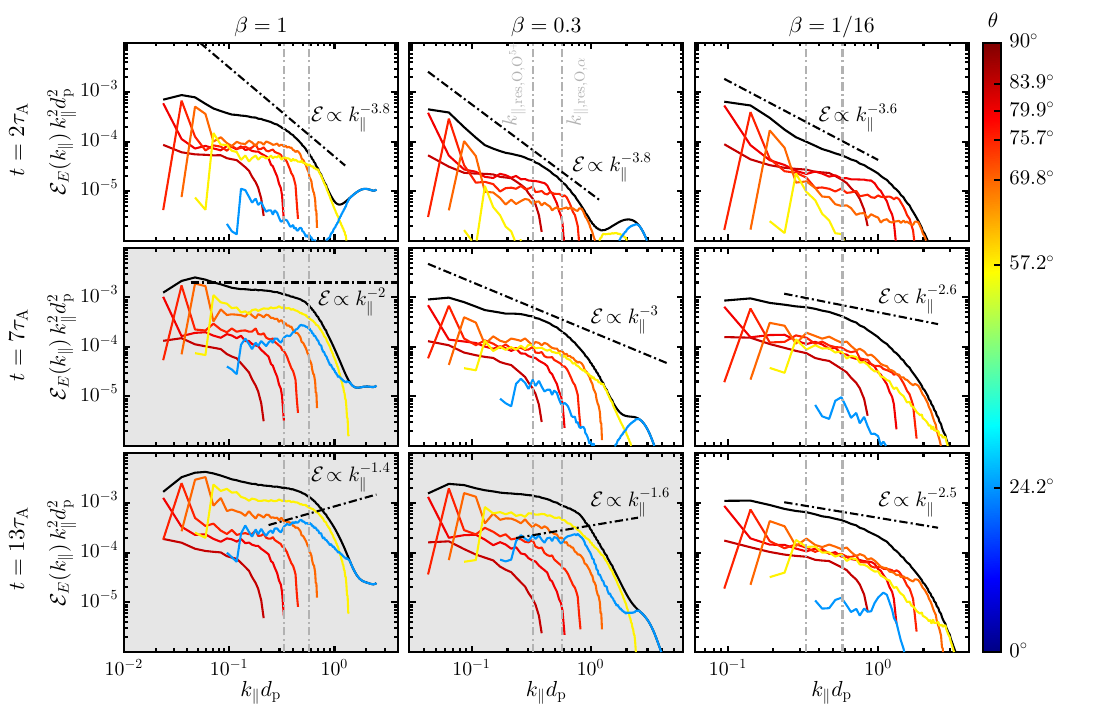}
    \caption{Electric power spectra in \(k_{\parallel }\), \(\mathcal{E}_{E_{\perp}}\) (black, solid), at \(t=2\tau _{\mathrm{A}} \)
      (top row), \(t=7\tau _{\mathrm{A}} \) (middle row), and \(t=13\tau _{\mathrm{A}} \) (bottom row) for the imbalanced
      simulations with  \(\beta_{\mathrm{p}0}  =1\) (left column),
    \(\beta _{\mathrm{p}0}= 0.3 \) (middle column), and
    \(\beta _{\mathrm{p}0} = 1 / 16 \) (right column). The spectra are computed from
    the spectra in \((k_{\perp},k_{\parallel})\) space in Fig.~\ref{fig:2dspec} by removing KAWs
  through a cutoff at \(k_{\perp } \rho _{\mathrm{p}0} = \pi/2   \) and integrating in
  \(k_{\perp } \). Different colors correspond to spectra evaluated over different
  equal-area bins in \((k_{\perp},k_{\parallel})\) specified by their central obliquity,
  \(\tan   \theta  = k_{\perp } / k_{\parallel }  \). Values of \(k_{\parallel } \), for which Alfv\'en waves are resonant
  with \({\mathrm{O}}^{5+} \) and alphas that have \(w_{\parallel } =0\), are indicated by
  the vertical, grey dash-dotted lines. Guidelines for various spectral slopes needed such that the measured values of \(a(t)\) in Fig.~\ref{fig:mqscaling} (given at each time in Table~\ref{tab:heating-ratios}) agree with the nondispersive prediction (\(a(t) = \eta -2 \) with \(\mathcal{E}_{E_{\perp}}\sim k_{\parallel}^{-\eta}\))
  are shown by black dash-dotted lines. Gray shading marks panels with appreciable power in the lowest-obliquity bin; these are the snapshots for which the heating-rate scaling with \(A_i/Z_i\) in Fig.~\ref{fig:mqscaling} is inverted, \(a<0\).}
    \label{fig:kprlspec}
\end{figure*}

To compare these empirically inferred power laws with the non-dispersive quasilinear prediction, \(a_{\mathrm{pred}}=\eta-2\) for \(\mathcal{E}_{E_{\perp}}\sim k_{\parallel}^{-\eta}\), we compute the power spectra of perpendicular electric-field fluctuations, \(\mathcal{E}_{E_{\perp}}=\mathcal{E}_{E_y}+\mathcal{E}_{E_z}\).
As for Fig.~\ref{fig:2dspec}, we first compute \(\mathcal{E}_{E_{\perp}}\) in \((k_{\perp},k_{\parallel})\) space and then integrate over \(k_{\perp}\) to obtain \(\mathcal{E}_{E_{\perp}}(k_{\parallel})\).
For the imbalanced simulations, the resulting \(\mathcal{E}_{E_{\perp}}(k_{\parallel})\) spectra are plotted as black lines in Fig.~\ref{fig:kprlspec} at three different times.
The spectra are compensated by \(k_{\parallel}^{2}\), so that a horizontal spectrum corresponds to \(\mathcal{E}_{E_{\perp}}\sim k_{\parallel}^{-2}\), the conservative-flux-cascade scaling for which our quasilinear theory predicts \(a=0\).
As discussed in \S\ref{sec:theory}, \(\mathcal{G}_{k_{\parallel,\mathrm{res},i}}f_i\) determines whether oblique or parallel PCWs dominate the heating, while the electric-field spectrum of the dominant waves determines the mass--charge scaling.
To separate contributions from modes with different obliquities, \(\tan\theta=k_{\perp}/k_{\parallel}\), we divide \((k_{\perp},k_{\parallel})\) space into equal-area bins in \(\theta\), whose right edges are indicated by the colored dashed lines in Fig.~\ref{fig:2dspec}.
Because we expect heating to be dominated by inertial-range fluctuations rather than by kinetic Alfv\'en waves (KAWs), we integrate \(\mathcal{E}_{E_{\perp}}(k_{\perp},k_{\parallel})\) over \(k_{\perp}\) within each bin only up to a cutoff at \(k_{\perp}\rho_{\mathrm{p}0}= \pi /2\).
The resulting binned spectra, \(\mathcal{E}_{E_{\perp}}(k_{\parallel})\), are shown by the colored lines in Fig.~\ref{fig:kprlspec}.\footnote{The steepness of the early-time spectra in the top row of Fig.~\ref{fig:kprlspec} is sensitive to the cutoff in \(k_{\perp}\rho_{\mathrm{p}0}\). Including a longer kinetic range flattens the spectra because of the contribution from KAWs. This effect is less prominent at later times, as in the middle and bottom rows of Fig.~\ref{fig:kprlspec}, because the helicity barrier causes inertial-range fluctuations to grow and dominate in amplitude over the KAWs below the steep transition range at \(k_{\perp}\rho_{\mathrm{p}0}\gtrsim1\).}
At later times, parallel PCWs appear as a bump in the \(k_{\parallel}\)-spectra of the lowest-obliquity bins near \(k_{\parallel,\mathrm{P}}^{\star}\approx k_{\parallel,\mathrm{cr}}^{\star}\).

We now compare the empirically fitted power-law indices from Fig.~\ref{fig:mqscaling}, \(a(t)\), with the phenomenology of \S\ref{sec:ql-minor-heat} for imbalanced turbulence.
We consider three representative stages: early times, \(t=2\tau_{\rm A}\), corresponding to \(a_{\max}\); intermediate times, \(t=7\tau_{\rm A}\), corresponding to \(a_{7\tau_{\rm A}}\); and late times, \(t=13\tau_{\rm A}\), corresponding to \(a_{\rm end}\).
In each panel of Fig.~\ref{fig:kprlspec}, the black dash-dotted line shows the spectral slope \(k_{\parallel}^{-[a(t)+2]}\) required for the non-dispersive quasilinear prediction, \(a_{\rm pred}=\eta-2\), to match the fitted \(a(t)\).
The relevant values of \(a(t)\) are taken from Fig.~\ref{fig:mqscaling} and summarized in Table~\ref{tab:heating-ratios}.
Because our quasilinear predictions compare minor-ion heating rates at \(w_{\parallel}=0\), the relevant spectral range in \(k_{\parallel}\) is bounded by the vertical dash-dotted gray lines in Fig.~\ref{fig:kprlspec}.
These lines mark the resonant \(k_{\parallel,{\rm res}}(w_{\parallel}=0)\) range for \(\mathrm{O}^{5+}\) and alphas, the minor ions with the largest and smallest mass-to-charge ratios, respectively, for resonance with oblique PCWs.
We organize the comparison into three phases of heating: (i) initial heating, represented by \(a_{\max}\); (ii) near-saturated oblique-PCW heating, represented by \(a_{7\tau_{\rm A}}\) for \(\beta_{\rm p0}=\{0.3,1/16\}\) and by \(a_{\rm end}\) for \(\beta_{\rm p0}=1/16\); and (iii) parallel-PCW-facilitated heating, represented by \(a_{7\tau_{\rm A}}\) for \(\beta_{\rm p0}=1\) and by \(a_{\rm end}\) for \(\beta_{\rm p0}=\{1,0.3\}\).

\begin{enumerate}
\renewcommand{\labelenumi}{(\roman{enumi})}
\item
At early times, \(t=2\tau_{\mathrm{A}}\), the spectra in the top row of Fig.~\ref{fig:kprlspec} correspond to the initial-heating regime described in \S\ref{sec:ql-minor-heat}.
In this regime, the quasilinear prediction assumes that all minor-ion VDFs are far from saturation, so that their velocity-space gradients scale with the initial thermal speeds of each species.
Over the resonant ranges in \(k_{\parallel}\), the spectral slopes are steeper than at later times and steep enough to produce the strong early-time scalings, \(a_{\max}\approx1.6{-}1.8\).
Because the dispersive correction in Eq.~\eqref{eq:ql-heat-rate-fullO} scales as \(\eta-3\), spectra with \(\eta>3\), as in this early stage, further favor heating of ions with larger \(A_i/Z_i\) and thereby relax the required spectral steepness.
The choice of \(w_{\parallel,\mathrm{res}}\), the \(k_{\perp}\) cutoff applied to the spectra, and this dispersive correction can modify the resonant \(k_{\parallel}\) range, measured spectral steepness, and the required slope, improving agreement between the theory and empirical scalings.
This is particularly relevant for the \(\beta_{\mathrm{p}0}=1/16\) case in the early-time stage.

\item
At intermediate and late times, the second regime of \S\ref{sec:ql-minor-heat} applies: ion VDFs are nearly quasilinearly flattened along oblique-PCW scattering contours, so their velocity-space gradients along those contours are similar between species.
When the energy in parallel PCWs remains negligible, oblique Alfv\'en/PCW fluctuations dominate the total \(\mathcal{E}_{E_{\perp}}(k_{\parallel})\) and are responsible for ion heating.
Because the cascade power is distributed over a range of obliquities, the oblique fluctuations do not all share the same scattering contours.
In a weaker analogue of the cross-contour transport by parallel PCWs described in \S\ref{sec:ql-focusing}, diffusion at one obliquity can therefore move particles across contours that are nearly saturated with respect to fluctuations at another obliquity.
Thus, heating by oblique PCWs, while weakened, persists in this near-saturated state.

Moving down the rows of Fig.~\ref{fig:kprlspec}, \(\mathcal{E}_{E_{\perp}}(k_{\parallel})\) flattens within the resonant range.
This spectral flattening is consistent with the weakening power-law dependence of minor-ion heating rates in Fig.~\ref{fig:mqscaling}.
When parallel PCWs remain weak and heating is solely by oblique fluctuations, the spectral slopes of the total electric-field power agree well with the empirical heating-rate scalings through \(a_{\rm pred}=\eta-2\).
For example, at \(t=7\tau_{\rm A}\), \(\eta\approx\{3,2.6\}\) and \(a_{7\tau_{\rm A}}=\{1,0.6\}\) for \(\beta_{\rm p0}=\{0.3,1/16\}\), while at \(t=13\tau_{\rm A}\), \(\eta\approx2.5\) and \(a_{\rm end}=0.5\) for \(\beta_{\rm p0}=1/16\).
For these slopes, which are only slightly steeper than the constant-flux value \(\eta=2\), dispersive corrections are small because they scale as \(\eta-3\).
For example, if \(\mathcal{E}_{E}\sim k_{\parallel}^{-2.5}\), dispersion changes the predicted mass--charge scaling only from \(Q_i\sim A_i(A_i/Z_i)^{0.5}\) to approximately \(Q_i\sim A_i(A_i/Z_i)^{0.4}\) over the range of \(\Omega_i\) considered here.

\item
A third heating stage arises once quasilinearly focused parallel PCWs contain sufficient energy near the minor-ion-resonant range of \(k_{\parallel}\).
This stage still belongs to the second regime of \S\ref{sec:ql-minor-heat}, in which ion VDFs are nearly flat along oblique contours and standard quasilinear heating by oblique fluctuations slows.
Parallel PCWs then strongly facilitate cross-contour diffusion, providing an alternative pathway for further minor-ion heating and becoming the relevant wave population for determining the mass--charge scaling.
This mechanism explains the subsequent inversion of the mass-to-charge dependence in the \(\beta_{\mathrm{p}0}=\{1,0.3\}\) imbalanced simulations.

Consider the shaded panels in Fig.~\ref{fig:kprlspec}: \(t=7\tau_{\mathrm{A}}\) and \(t=13\tau_{\mathrm{A}}\) for \(\beta_{\mathrm{p}0}=1\), and \(t=13\tau_{\mathrm{A}}\) for \(\beta_{\mathrm{p}0}=0.3\).
At these times, quasilinear predictions based on the \(k_{\parallel}\)-slope of the total \(\mathcal{E}_{E_{\perp}}\) substantially over-predict the measured \(a\) in Fig.~\ref{fig:mqscaling}.
Concurrently, the lowest-obliquity blue bin contains relatively more power than in the other panels.
Because the parallel PCWs are driven by protons and peak at a finite \(k_{\parallel,\mathrm{P}}^{\star}\), at smaller parallel scales than those resonant with most minor ions, they flatten the \(k_{\parallel}\)-spectrum of the lowest-obliquity bin across the minor-ion-resonant range.
When this spectrum becomes shallower than the constant-flux scaling, \(k_{\parallel}^{-2}\), the predicted heating-rate scaling inverts, \(Q_i\sim A_i(A_i/Z_i)^a\) with \(a_{\mathrm{pred}}=\eta-2<0\).
This matches the inversions of the measured \(a\) in Fig.~\ref{fig:mqscaling} for the \(\beta_{\mathrm{p}0}=0.3\) and \(\beta_{\mathrm{p}0}=1\) imbalanced simulations.
These parallel PCWs reach large amplitudes earlier and at lower \(k_{\parallel}d_{\mathrm{p}0}\) for higher plasma beta.
This trend is consistent with the heating-rate inversion occurring by \(7\tau_{\mathrm{A}}\) for \(\beta_{\mathrm{p}0}=1\), by \(12\tau_{\mathrm{A}}\) for \(\beta_{\mathrm{p}0}=0.3\), and not occurring for \(\beta_{\mathrm{p}0}=1/16\).
\end{enumerate}

In the balanced-turbulence simulations, we do not observe a large population of parallel PCWs or an inversion of the heating-rate dependence on \(A_i/Z_i\).
The fitted scalings from these simulations are shown in red in the bottom panel of Fig.~\ref{fig:mqscaling}.
Their evolution is less well captured by quasilinear predictions based on the evolving \(k_{\parallel}\)-slopes of the electric-field spectra.
The spectra for the balanced runs are not shown.
This weaker agreement is expected due to stronger nonlinear frequency broadening in balanced turbulence \citep{johnstonQuasilinearTheoryPerpendicular2025}.
We therefore view the balanced simulations primarily as a control case demonstrating that the inversion mechanism is promoted in imbalanced turbulence, where quasilinear focusing is enhanced by the helicity barrier.

Putting these results together, the saturated minor-ion heating rates in our simulations can be expressed by the semi-empirical formula
\begin{equation}
\label{eq:emp-form}
\frac{Q_i}{Q_{\rm p}}
\approx
A_i\left(\frac{A_i}{Z_i}\right)^a \times \begin{cases}
1.0, & \text{balanced}\\
1.7, & \text{imbalanced}
\end{cases} ,
\end{equation}
where we have neglected the weak \(\beta_{\mathrm{p}0}\) dependence of the coefficient and the dispersive corrections described in \S\ref{sec:ql-minor-heat}.
For imbalanced turbulence, the helicity barrier predicts an electron heating rate of \(Q_{\rm e}/(\mathcal{V}\varepsilon_{\rm inj})\approx1-\varepsilon_{\rm inj,H}/\varepsilon_{\rm inj}\), which then implies in saturation that the proton heating rate is \(Q_{\rm p}/(\mathcal{V}\varepsilon_{\rm inj})\approx\varepsilon_{\rm inj,H}/\varepsilon_{\rm inj}\).
The mass--charge dependence of \(Q_i\) is then set by the quasilinear phenomenology of \S\ref{sec:ql-minor-heat}, with \(a\approx\eta-2\) determined by the electric-field spectrum, \(\mathcal{E}_{E_{\perp}}\sim k_{\parallel}^{-\eta}\), of the wave population that controls the heating.
When parallel PCWs control the heating, their flatter spectra can invert the dependence, causing heating rates to correlate negatively with mass-to-charge ratio.

With this formula, we can contextualize the trends of \(T_i/T_{i0}\) with imbalance and \(\beta_{\mathrm{p}0}\) from \S\ref{sec:time-evol-turb}.
Despite the lower values of \(\varepsilon_{\rm inj}\) in the imbalanced runs (Table~\ref{tab:results-summary}), minor-ion temperatures become most extreme when turbulence is imbalanced because both proton heating relative to electron heating, \(Q_{\rm p}/Q_{\rm e}\), and minor-ion heating relative to proton heating, \(Q_i/Q_{\rm p}\), are enhanced.
The helicity barrier gives rise to both enhancements, while the dispersive properties of PCWs, which affect protons most strongly, also contribute to the enhancement of \(Q_i/Q_{\rm p}\).
Although \(\varepsilon_{\rm inj}\), and therefore \(Q_i\), varies only weakly with \(\beta_{\mathrm{p}0}\) in the imbalanced runs, the fractional temperature increase, \(T_i/T_{i0}\), grows strongly toward lower \(\beta_{\mathrm{p}0}\).
This trend follows from our choice to normalize the forcing to the magnetic-field energy, or equivalently to comparable Alfv\'enic fluctuation amplitudes, rather than to the initial thermal energy: for comparable \( u_{\perp}/v_{\rm A}\), the fluctuation energy represents a larger fraction of the initial ion thermal energy at lower beta.
Together with the tendency for \(a(t)\) to saturate at higher values at lower \(\beta_{\mathrm{p}0}\), owing to weaker parallel-PCW growth, this explains why minor-ion temperatures are most extreme in low-\(\beta_{\mathrm{p}0}\), imbalanced turbulence.

\section{Summary}
\label{sec:summary}

As tracers, minor ions provide rich diagnostics of the kinetic physics governing waves, turbulence, and dissipation throughout the collisionless solar wind.
Their wide range of masses and charge states tightly constrains theories that seek to explain their extreme perpendicular temperatures and the dependence of those temperatures on mass-to-charge ratio, \(A_i/Z_i\).
We develop a quasilinear phenomenology for this mass-to-charge dependence, predicting how minor-ion heating rates scale with \(A_i/Z_i\) as Alfv\'enic turbulence and heating evolve toward saturation.
Combined with the physics of turbulence imbalance and the helicity barrier, this phenomenology predicts the salient features of minor-ion heating in hybrid-kinetic simulations across solar-wind-relevant parameters: proton beta, \(\beta_{\mathrm{p}0}\in\{1,0.3,1/16\}\), and normalized cross helicity, \(\sigma_{\mathrm{c}}\approx\{0,0.9\}\).

Minor-ion temperatures and temperature anisotropies become most extreme and most perpendicular in imbalanced turbulence at low \(\beta_{\mathrm{p}0}\).
For example, by the end of our imbalanced \(\beta_{\mathrm{p}0}=1/16\) simulation, we find \(T_{\perp\mathrm{O}^{5+}}/T_{\perp\mathrm{p}}\approx39\) and \(T_{\perp\mathrm{O}^{5+}}/T_{\parallel\mathrm{O}^{5+}}\approx10\), reproducing remote UVCS measurements of coronal holes \citep{cranmerImprovedConstraintsPreferential2008,kohlUltravioletSpectroscopyExtended2006}.
The extreme temperatures arise from enhanced proton heating relative to electron heating, \(Q_{\rm p}/Q_{\rm e}\), and enhanced minor-ion heating relative to proton heating, \(Q_i/Q_{\rm p}\propto K/Q_{\rm p}\), when turbulence is imbalanced.
In our simulations, \(K\approx1.7 Q_{\rm p}\) for imbalanced turbulence and \(K\approx1.0 Q_{\rm p}\) for balanced turbulence.
Both enhancements arise from the helicity barrier, while the enhancement of \(Q_i/Q_{\rm p}\) is further strengthened by stronger dispersive effects for resonant protons.
At lower \(\beta_{\mathrm{p}0}\), the fluctuation energy is also larger relative to the initial ion thermal energy.
Likewise, the temperature-anisotropy trends follow the VDF shapes predicted by quasilinear saturation in imbalanced turbulence.

Our phenomenology predicts that minor-ion heating rates scale as \(Q_i \approx K A_i(A_i/Z_i)^a(v_{\mathrm{ph},k_{\parallel,\mathrm{res}}}^{\eta}/v_{{\rm g},k_{\parallel,\mathrm{res}}})\).
Here, \(a=\eta-2\), where \(\eta\) is the \(k_{\parallel}\)-spectral slope of the perpendicular-electric-field spectrum, \(\mathcal{E}_{E_{\perp}}\sim k_{\parallel}^{-\eta}\), over the minor-ion-resonant range of the wave population responsible for heating.
Unlike previous quasilinear treatments of the mass--charge scaling of minor-ion heating \citep{isenbergPreferentialAccelerationHeating1983a,hollwegCyclotronResonanceCoronal1999,cranmerCoronalHolesHighSpeed2002}, our model retains both oblique and parallel PCWs.
Their relative importance changes as ion VDFs approach quasilinear saturation along oblique-PCW contours, and both wave populations emerge self-consistently in imbalanced turbulence.
Because of the helicity barrier, oblique Alfv\'enic fluctuations in the inertial range develop smaller parallel scales and become dispersive, oblique PCWs.
The resulting strong cyclotron-resonant proton heating by these high-frequency oblique PCWs drives the proton population unstable to the emission of parallel PCWs through \emph{quasilinear focusing} \citep{chandranResonantInteractionsProtons2010}.
When minor-ion VDFs are nearly quasilinearly saturated along oblique-PCW scattering contours, these parallel PCWs facilitate additional cross-contour diffusion and enable further heating by oblique fluctuations. Unlike oblique fluctuations, whose energy decreases as they cascade to higher \(k_{\parallel}\), these parallel PCWs concentrate power near large \(k_{\parallel}\sim k_{\parallel,\mathrm{P}}^{\star}\).
Because minor ions resonate strongly at \(k_{\parallel,{\rm res},i}\sim \Omega_i / v_{\rm A}\), and the power in parallel PCWs decreases away from \(k_{\parallel,\mathrm{P}}^{\star}>k_{\parallel,{\rm res},i}\), minor ions with higher gyrofrequency \(\Omega_i \propto Z_i / A_i \) (i.e., smaller mass-to-charge ratio \(A_i / Z_i\)) become preferentially heated.
This process can invert the mass-to-charge scaling of \(Q_i\), changing \(a\) from positive values to \(a<0\).

We measure \(a(t)\) for minor-ion heating rates in our simulations and find that its evolution is consistent with our phenomenology when the turbulence is imbalanced.
At early times, before parallel PCWs develop appreciable power, the oblique-dominated spectra are steep, approximately \(k_{\parallel}^{-4}\), and the power-law scaling peaks near \(a\approx2\).
As the turbulence evolves, the spectra flatten, accompanied by a corresponding decrease in \(a(t)\).
When parallel PCWs reach sufficient amplitude, the predicted inversions occur, with \(a<0\).
In contrast, our simulations with balanced turbulence exhibit weaker, less perpendicular minor-ion heating and no inversion of the heating-rate dependence on \(A_i/Z_i\).
Because nonlinear broadening is strong in balanced turbulence \citep{johnstonQuasilinearTheoryPerpendicular2025}, our quasilinear phenomenology is less applicable in this regime and the balanced simulations do not produce parallel PCWs for the same reasons.

We therefore predict that, in the solar wind, ``inverted'' scalings of minor-ion temperature with ion mass-to-charge ratio,
\(T_i/A_i\propto(A_i/Z_i)^a\) with \(a<0\), should preferentially occur in intervals with a history of enhanced coherent parallel PCW power and large cross helicity.
A caveat is that observed minor-ion temperatures reflect the accumulated history of heating along the flow, whereas \(a(t)\) in our simulations describes the instantaneous dependence of heating rates on the local turbulent spectra.
Nevertheless, if the regions that contribute most strongly to minor-ion temperatures are also highly imbalanced, then ion-cyclotron heating, quasilinear focusing, and parallel-PCW emission should operate efficiently there.
In that case, the connection between inverted temperature scalings, parallel PCWs, and cross helicity may remain visible in the accumulated minor-ion temperature signatures.

Previous observational studies of minor-ion temperatures, including inversions with mass-to-charge ratio \citep{tracyConstrainingSolarWind2016}, have often organized intervals by collisional age.
For ion--proton thermalization, the species-dependent collisional age can be written approximately as \(A_{{\rm C},i}\sim (R/v_{\rm sw})\nu_{i{\rm p}}\), where \(\nu_{i{\rm p}}\) is the Coulomb thermalization rate between ion species \(i\) and protons \citep{kasperHotSolarWindHelium2008}.
Thus, low-\(A_{\rm C}\) intervals tend to correspond to faster, less collisionally processed solar wind, and \citet{tracyConstrainingSolarWind2016} find broader ranges of inverted thermal-speed scalings in such intervals.
Because solar-wind speed, collisional age, and cross helicity are themselves correlated, ongoing measurements from \textit{Parker Solar Probe} and \textit{Solar Orbiter} should seek to also isolate cross helicity as an independent organizing parameter for minor-ion heating.
The most direct tests are whether increasing cross helicity correlates with stronger coherent parallel PCWs, enhanced minor-ion temperatures, larger perpendicular temperature anisotropies, and more frequent or stronger inverted scalings with mass-to-charge ratio.
If these correlations are observed, they would provide direct evidence that turbulence imbalance and the helicity barrier govern particle heating in high-cross-helicity regions of the solar wind.

\vspace{1cm}
\noindent This work benefited from useful conversations with Kristopher Klein, Benjamin Chandran,
Christopher Chen, Mihailo Martinovi\'c, Romain Meyrand, and Zade Johnston. M.F.Z.,  M.W.K., and E.Y.~were
supported by the National Aeronautics and Space Administration (NASA) under
Grant No.~80NSSC24K0171 issued through the Heliophysics, Theory, Modeling and
Simulation Program. Support for J.S.~was provided by Rutherford Discovery
Fellowship RDF-U001804 and Marsden Fund grant UOO1727, which are managed through
the Royal Society Te Ap\=arangi. Additional support for E.Y.~was provided by NASA grant No.~NNN06AA01C. This
work is part of the Frontera computing project at the Texas Advanced Computing
Center under allocation number AST20010; it also made extensive use of the
Stellar cluster at the PICSciE-OIT TIGRESS High Performance Computing Center and
Visualization Laboratory at Princeton University. The authors thank the Kavli Institute for Theoretical Physics (KITP)
for its hospitality during the completion of this work; KITP is supported in
part by the National Science Foundation under Award No.~PHY-2309135.

\end{document}